\documentclass[%
 reprint,
 amsmath,amssymb,
 aps,
]{revtex4-2}
\pdfoutput=1
\usepackage[utf8]{inputenc}
\usepackage[english]{babel}
\usepackage[T1]{fontenc}
\usepackage{amsmath}
\usepackage{amssymb}
\usepackage{amsfonts}
\usepackage{hyperref}
\usepackage{cellspace}
\usepackage{mathtools}
\usepackage{cleveref}
\usepackage[export]{adjustbox}
\usepackage[caption=false]{subfig}
\usepackage{xcolor}

\usepackage{xparse}

\newcommand{\hyp}{\mathbb{H}^2}
\newcommand{\pslR}{\operatorname{PSL_2(\mathbb{R})} \rtimes \mathbb{Z}_2}

\begin{document}

\title{Constructions and performance of hyperbolic and semi-hyperbolic Floquet codes}

\author{Oscar Higgott}
\email{oscar.higgott.18@ucl.ac.uk}
\affiliation{Department of Physics \& Astronomy, University College London, WC1E 6BT London, United Kingdom}
\author{Nikolas P.~Breuckmann}
\email{niko.breuckmann@bristol.ac.uk}
\affiliation{School of Mathematics, University of Bristol, Fry Building, Woodland Road, Bristol BS8 1UG, United Kingdom}

\begin{abstract}
We construct families of Floquet codes derived from colour code tilings of closed hyperbolic surfaces.
These codes have weight-two check operators, a finite encoding rate and can be decoded efficiently with minimum-weight perfect matching.
We also construct semi-hyperbolic Floquet codes, which have improved distance scaling, and are obtained via a fine-graining procedure.
Using a circuit-based noise model that assumes direct two-qubit measurements, we show that semi-hyperbolic Floquet codes can be $48\times$ more efficient than planar honeycomb codes and therefore over $100\times$ more efficient than alternative compilations of the surface code to two-qubit measurements, even at physical error rates of $0.3\%$ to $1\%$.
We further demonstrate that semi-hyperbolic Floquet codes can have a teraquop footprint of only 32 physical qubits per logical qubit at a noise strength of $0.1\%$.
For standard circuit-level depolarising noise at $p=0.1\%$, we find a $30\times$ improvement over planar honeycomb codes and a $5.6\times$ improvement over surface codes.
Finally, we analyse small instances that are amenable to near-term experiments, including a Floquet code derived from the Bolza surface that encodes four logical qubits into 16 physical qubits.
\end{abstract}

\maketitle

\section{Introduction}

Recent breakthroughs in the development of quantum LDPC codes have led to significant improvements in the asymptotic parameters of quantum codes~\cite{breuckmann2021ldpc, hastings2021fiber,panteleev2022quantum, breuckmann2021balanced, panteleev2022asymptotically,leverrier2022quantum}.
However, demonstrating a significant advantage over the surface code for realistic noise models and reasonable system sizes can be challenging.
For example, Ref.~\cite{tremblay2022constant} demonstrated a significant $14\times$ saving in resources relative to the surface code using hypergraph product codes for circuit-level noise, however this improvement was shown for very low physical error rates of around $0.01\%$ and for very large system sizes (around 13 million qubits).
This is in large part due to the high-weight stabiliser generators common in many families of quantum LDPC codes, resulting in deep and complex syndrome extraction circuits, which can introduce hook errors and lower their noise thresholds~\cite{conrad2018small,li2020numerical,tremblay2022constant,pattison2023hierarchical}.

Since we ultimately need to construct a fault-tolerant circuit implementing a code, it is therefore highly desirable to use a code that supports low-weight check operators.
A broad family of codes that can have this property are subsystem codes~\cite{poulin2005stabilizer}.
Subsystem codes are a generalisation of stabiliser codes that can allow high-weight stabilisers to be measured by taking the product of measurement outcomes of low-weight anti-commuting check operators (gauge operators).
Several families of subsystem codes with low-weight check operators have been discovered~\cite{bacon2006operator,suchara2011constructions,bravyi2013subsystem,kubica2022single}, and in Ref.~\cite{higgott2021subsystem} we constructed finite-rate subsystem codes with weight-three check operators, derived from tessellations of closed hyperbolic surfaces.
The low-weight checks and symmetries of the subsystem hyperbolic codes in Ref.~\cite{higgott2021subsystem} led to efficient syndrome extraction circuits, which are $4.3\times$ more efficient than surface code circuits for $0.1\%$ circuit-level noise and for a system size of around 19 thousand physical qubits.
However, it is natural to ask if we can reduce the check weight further and demonstrate larger resource savings relative to the surface code.

\begin{figure}[t]
    \centering
    \includegraphics[width=0.8\linewidth]{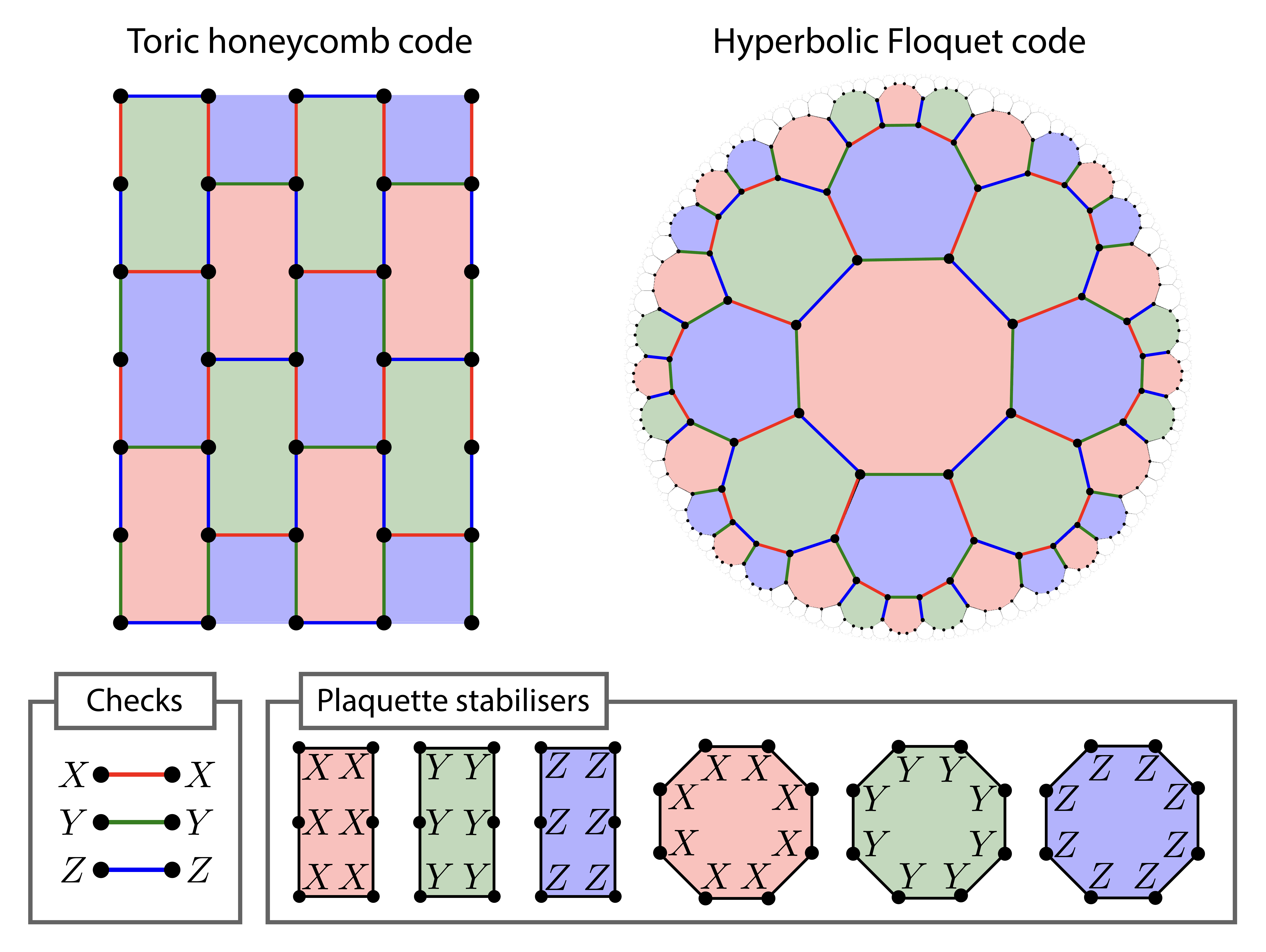}
    \caption{Two of the families of Floquet codes studied in this work. A data qubit is associated with each vertex of a lattice, a two-qubit check operator is associated with each edge and a plaquette stabiliser generator is associated with each face.  Left: A toric honeycomb code, which is defined on a hexagonal lattice with periodic boundary conditions and three-colourable faces~\cite{Hastings2021dynamically}. Here we consider a patch with 4 columns of data qubits and 6 rows (dimensions $4\times 6$). Opposite sides of the lattice are identified.
    Right: A Floquet code derived from a tiling of a hyperbolic surface with three octogans meeting at each vertex (vertex configuration 8.8.8) and three-colourable faces. Here we draw a region of the tiled hyperbolic plane defining the code family, however each Floquet code in the family is derived from a tiling of a closed surface.}
    \label{fig:honeycomb_and_hyperbolic_floquet_checks}
\end{figure}

We achieve further improvements in this work by constructing codes from the broader family of Floquet codes~\cite{Hastings2021dynamically}, which can be seen as a generalisation of subsystem codes, see e.g.~Ref.~\cite{townsendteague2023floquetifying} for a discussion.
Similar to subsystem codes, Floquet codes involve the measurement of low-weight anti-commuting check operators.
Unlike subsystem codes, however, Floquet codes do not require that the logical operators of the code remain static over time.
Indeed, Floquet codes do not admit a static set of generators for all their logical Pauli operators, and the form of these logical operators instead evolves periodically during the syndrome extraction circuit.
This additional flexibility leads to codes with weight-two check operators and good performance~\cite{gidney2021fault}, especially in platforms that support direct two-qubit Pauli measurements, such as Majorana-based qubits~\cite{paetznick2023performance}.
Notably, the planar honeycomb code~\cite{haah2022boundaries,paetznick2023performance,gidney2022benchmarking} is a specific Floquet code derived from a hexagonal lattice with open boundary conditions, and is therefore amenable to experimental realisation on a quantum computer chip (e.g.~a solid state device such as a superconducting qubit architecture).

In Ref.~\cite{vuillot2021planar}, Vuillot showed how Floquet codes can be derived from any tiling suitable to define a colour code~\cite{bombin2006topological}, and furthermore observed that Floquet codes derived from colour code tilings of closed hyperbolic surfaces, which we refer to as \textit{hyperbolic Floquet codes}, would have a constant encoding rate and logarithmic distance.
However, there has not been prior work constructing explicit examples of hyperbolic Floquet codes or analysing their performance.

In this work, we construct explicit families of hyperbolic Floquet codes (see \Cref{fig:honeycomb_and_hyperbolic_floquet_checks}) and analyse their performance numerically, comparing them to planar honeycomb codes and surface codes.
We also construct Floquet codes derived from semi-hyperbolic lattices, which fine-grain hyperbolic lattices, leading to an improved distance scaling that enables exponential suppression of errors while retaining an advantage over honeycomb and surface codes.
Furthermore, the thresholds of families of semi-hyperbolic Floquet codes are essentially the same as the threshold of the honeycomb code.
All of our constructions have weight-two check operators and can be decoded efficiently using surface code decoders, such as minimum-weight perfect matching~\cite{dennis2002topological,fowler2014minimum,higgott2023sparse,wu2023fusion}.

For platforms that support direct two-qubit measurements, we find that semi-hyperbolic Floquet codes can require $48\times$ fewer physical qubits than honeycomb codes even for high physical error rates of 0.3\% to 1\% (for the `EM3' noise model) and for a system size of 21,504 physical qubits.
These results imply that our constructions are over $100\times$ more efficient than alternative compilations of the surface code to two-qubit measurements~\cite{Chao2020optimizationof,gidney2022pair}, which have been shown to be less efficient than honeycomb codes for the same noise model~\cite{paetznick2023performance,gidney2022pair}.
We also show that semi-hyperbolic Floquet codes can require as few as 32 physical qubits per logical qubit to achieve logical failure rates below $10^{-12}$ per logical qubit at a physical error rate of $0.1\%$, whereas honeycomb codes instead require 600~\cite{gidney2021fault} to 2000~\cite{paetznick2023performance} physical qubits per logical qubit in the same regime.

We also consider a standard circuit-level depolarising noise model (``SD6''), for which two-qubit measurements are implemented using an ancilla, CNOT gates and single-qubit rotations.
For this noise model, our semi-hyperbolic Floquet codes are $30\times$ more efficient than planar honeycomb codes and over $5.6\times$ more efficient than conventional surface codes for physical error rates of around $0.1\%$ and below.

Finally, we construct small examples of hyperbolic Floquet codes that are amenable to near-term experiments.
This includes codes derived from the Bolza surface, using as few as 16 physical qubits, and which we show are $3\times$ to $6\times$ more efficient than their Euclidean counterparts.

Although these (semi-)hyperbolic Floquet codes cannot be implemented using geometrically local connections in a planar Euclidean architecture, we show how they can instead be implemented using a \textit{biplanar} or \textit{modular} architecture.
A biplanar architecture uses two layers of qubits, where connections within each layer do not cross, but may be long-range.
A modular architecture consists of many small modules, where each module only requires local 2D planar Euclidean connectivity and connections between modules may be long-range.
The long-range connections between modules could be mediated via photonic links in a trapped-ion architecture~\cite{monroe2014large}, for example.

 We note that some related works on LDPC codes were posted to arXiv shortly after the first version of our manuscript. 
 Refs.~\cite{Bravyi2024highthreshold} and \cite{Xu2024constant} demonstrate a $\approx 10\times$ reduction in qubit overhead relative to the surface code for circuit-level noise of around 0.1\% using bivariate bicycle (BB) and lifted product (LP) codes, respectively, and Ref.~\cite{fahimniya2023fault} also studied hyperbolic Floquet codes.
 
Our work complements Refs.~\cite{Bravyi2024highthreshold,Xu2024constant}, since our hyperbolic Floquet codes excel for the EM3 noise model (pair measurement gate set) that is likely challenging for the BB and LP codes.
Specifically, we achieve a $\sim10\times$ to $20\times$ qubit reduction relative to surface codes for the EM3 noise model using a similar system size to Refs.~\cite{Bravyi2024highthreshold,Xu2024constant} (a few hundred qubits), with the improvement rising to a $\sim 100\times$ reduction in qubit overhead for larger system sizes.
Given the check weight of the BB and LP codes it seems unlikely that a compilation to pair measurements would perform well, although this has not been explored.
On the other hand, the codes in Refs.~\cite{Bravyi2024highthreshold,Xu2024constant} perform better for a standard circuit-level noise model, for which we obtain a $\sim 2.6\times$ qubit saving relative to the surface code using similar system sizes (hundreds of qubits) as those used for their their $\sim 10 \times$ saving. 
Our improvement rises to a $5.6\times$ saving for larger system sizes.
Furthermore, unlike Refs.~\cite{Bravyi2024highthreshold,Xu2024constant}, our codes are amenable to fast decoders based on matching and our connectivity graph has a lower degree of qubit connectivity (degree three, compared to 6 or more in~\cite{Bravyi2024highthreshold,Xu2024constant}).

Our work also complements Ref.~\cite{fahimniya2023fault}, which also studies hyperbolic Floquet codes.
Both our work and theirs considers 6.6.6 tilings, however we find some instances not considered in Ref.~\cite{fahimniya2023fault}, and likewise they construct some 6.6.6 tilings not considered in our work.
Additionally, our work also considers 4.10.10 and 4.8.10 tilings as well as semi-hyperbolic tilings, which are necessary to suppress errors exponentially and obtain reasonable teraquop footprint estimates.

Our paper is organised as follows.
We start by introducing Euclidean, hyperbolic and semi-hyperbolic colour code tilings in \Cref{sec:colour_code_tilings}.
In \Cref{sec:floquet_codes} we review Floquet codes, including how we construct Floquet code circuits from the colour code tilings of \Cref{sec:colour_code_tilings}, as well as how these circuits can be decoded.
\Cref{sec:floquet_codes} also describes the EM3 and SD6 noise models we use in our simulations.
In \Cref{sec:constructions} we present an analysis of the (semi-)hyperbolic Floquet codes we have constructed, including a study of their parameters (\Cref{sec:code_parameters}) and simulations comparing them to honeycomb and surface codes (\Cref{sec:simulations}).
We then conclude in \Cref{sec:conclusion} with a summary of our results and a discussion of future work.

\section{Colour code tilings}
\label{sec:colour_code_tilings}

Floquet codes~\cite{Hastings2021dynamically} can be defined on any tiling of a surface that is also suitable to define a 2D colour code~\cite{bombin2006topological}; namely, it is sufficient that the faces are 3-colourable and the vertices are 3-valent~\cite{vuillot2021planar}.
We will refer to such a tiling as a \textit{colour code tiling}.
In this section we will review colour code tilings and describe the tilings used to construct Floquet codes in this work.

We denote by $\mathcal{T}=(V, E, F)$ a colour code tiling with vertices $V$, edges $E\subset V^2$ and faces $F\subset 2^V$.
Each face $f\in F$ is assigned a colour $C(f)\in\{R, G, B\}$ which can be red, green or blue ($R$, $G$ or $B$), such that two faces that share an edge have different colours:
\begin{equation}
    f_1\cap f_2 \in E \implies C(f_1)\neq C(f_2),\quad\forall (f_1, f_2) \in F^2.
\end{equation}
We will also assign a colour $C(e)\in\{R,G,B\}$ to each edge $e$, given by the colour of the faces that it links.
Equivalently, the colour of an edge is the complement of the colours of the two faces it borders.
See \Cref{fig:honeycomb_and_hyperbolic_floquet_checks} for two examples of colour code tilings.

We now introduce some definitions from homology which are applicable to any tiling $\mathcal{T}=(V,E,F)$ of a closed surface, including tilings that are not colour code tilings.
The \textit{dual} of $\mathcal{T}$, which we denote $\mathcal{T}^*=(V^*,E^*,F^*)$, has a vertex for each face of $\mathcal{T}$ and two vertices in $\mathcal{T}^*$ are connected by an edge if the corresponding faces in $\mathcal{T}$ share an edge.
A \textit{cycle} in $\mathcal{T}$ is a set of edges (a subset of $E$) that forms a collection of closed paths in~$\mathcal{T}$ (it has no boundary).
A \textit{co-cycle} in $\mathcal{T}$ is a set of edges (also a subset of $E$) that corresponds to a cycle in $\mathcal{T}^*$.
A cycle is \textit{homologically non-trivial} if it is not the boundary of a set of faces in $\mathcal{T}$ and is homologically trivial otherwise.
Likewise, a co-cycle is homologically non-trivial if it does not correspond to the boundary of a set of faces in $\mathcal{T}^*$ and is homologically trivial otherwise.
Two cycles belong to the same homology class if the symmetric difference of the two cycles is a homologically trivial cycle and the dimension of the first homology group of the tiling, $\dim H_1$, is the number of distinct homology classes.

\subsection{Hyperbolic surfaces with colour code tilings}\label{sec:hyperbolic_colour_code_tilings}
Closed hyperbolic surfaces can be obtained via a compactification procedure.
Consider the infinite hyperbolic plane $\hyp$, which is the universal covering space of hyperbolic surfaces, and its group of isometries $\pslR$.
Taking a subgroup $\Gamma \subset \pslR$ which acts fixed-point free and discontinuous on $\hyp$, we obtain a closed hyperbolic surface by taking the quotient $X = \hyp/\Gamma$.
Note that due to the Gau\ss --Bonnet theorem, the area of the surface $\operatorname{area}(X)$ is proportional to the dimension of the first homology group $\dim H_1(X)$.

This idea extends to tiled surfaces:
consider a periodic tiling on~$\hyp$ and a group~$\Gamma$ which preserves the tiling structure, then the quotient surface $X = \hyp/\Gamma$ supports the same tiling.
In particular, if we consider a periodic colour code tiling of $\hyp$ and a group~$\Gamma \subset \pslR$ that is compatible with it, we obtain a closed hyperbolic surface with the desired tiling.

\subsection{Wythoff construction}
\label{sec:wythoff}

\begin{figure}
    \centering
    \includegraphics[width=0.7\linewidth]{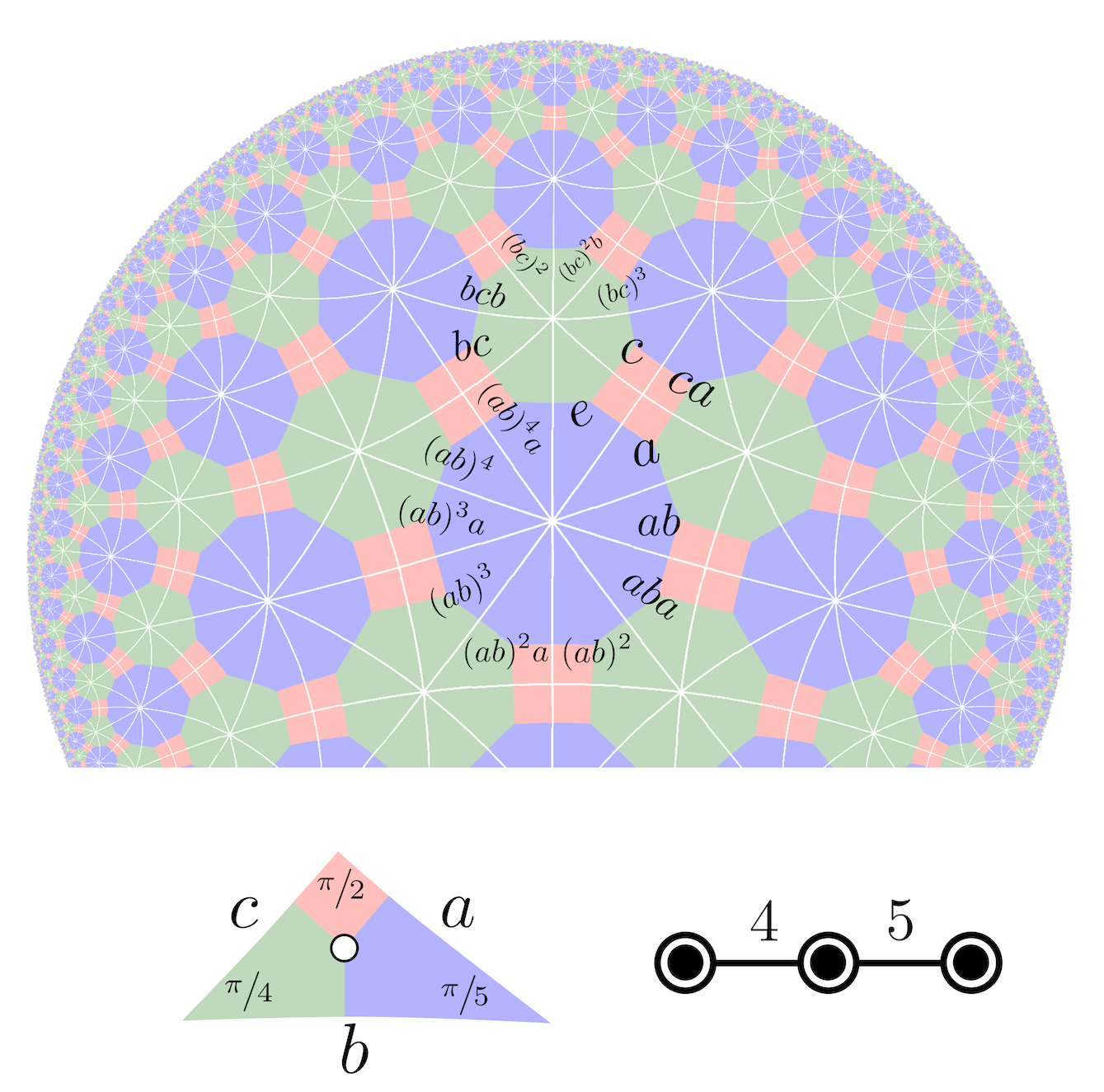}
    \caption[Wythoff's kaleidoscopic construction for the triangle group $\Delta(2,4,5)$]{A hyperbolic colour code tiling generated by the hyperbolic triangle group $\Delta(2,4,5)$ using Wythoff's kaleidoscopic construction. The tiling is generated through reflections across the sides of a fundamental triangle with internal angles $\pi/2$, $\pi/4$ and $\pi/5$. We show a white cut along each reflection line. Below the tiling we show the fundamental triangle and Coxeter diagram. A generator point is placed in the interior of the fundamental triangle and from each mirror a perpendicular line is drawn to the generator point. This generates a uniform tiling with vertex configuration $4.8.10$ and three-colourable faces.}
    \label{fig:wythoff_generator_points}
\end{figure}

We can obtain suitable colour code tilings of the hyperbolic plane using Wythoff's kaleidoscopic construction.
This approach has been used in \cite{vuillot2022quantum} to define hyperbolic colour codes and pin codes.
We now review this method and explain how we use it to generate hyperbolic colour code tilings.

We choose a \textit{fundamental triangle} with internal angles $\pi/p$, $\pi/q$ and $\pi/t$.
We can generate an infinite tiling through reflections in the sides of the triangle (we refer to each side of the triangle as a mirror).
The group generated by these reflections (a symmetry group of the tiling) is the triangle group $\Delta(p,q,t)$, an example of a Coxeter group.
We place a vertex, called a generator point, within the fundamental triangle, and from each mirror we draw a perpendicular line to the generator point.
These lines are mapped by the reflections to edges in the tiling, and the generator point is similarly mapped to vertices of the tiling.
For a given triangle group $\Delta(p,q,t)$ we can generate different tilings depending on the choice of generator point; in this work we place the generator point in the interior of the fundamental triangle to generate colour code tilings, as shown shown in \Cref{fig:wythoff_generator_points} for $\Delta(2,4,5)$.

If we denote a reflection in each of the three sides by $a$, $b$ and $c$, where the angles between the pairs of mirrors $(a,b)$, $(b,c)$ and $(c,a)$ are $\pi/t$, $\pi/q$ and $\pi/p$ respectively, then the triangle group $\Delta(p,q,t)$ has presentation
\begin{multline}
\Delta(p,q,t)=\langle a,b,c~|~a^2=b^2=c^2=(ab)^t\\=(bc)^q=(ca)^p=e \rangle
\end{multline}
where $e$ is the identity element.
The relations for $(ab)^t$, $(bc)^q$ and $(ca)^p$ can be understood from the fact that $ab$, $bc$ and $ca$ are rotations by $2\pi/t$, $2\pi/q$ and $2\pi/p$, respectively.
We note that if $1/p+1/q+1/t < 1$ then we obtain a colour code tiling of the hyperbolic plane.

\begin{figure}
    \centering
    \includegraphics[width=0.7\linewidth]{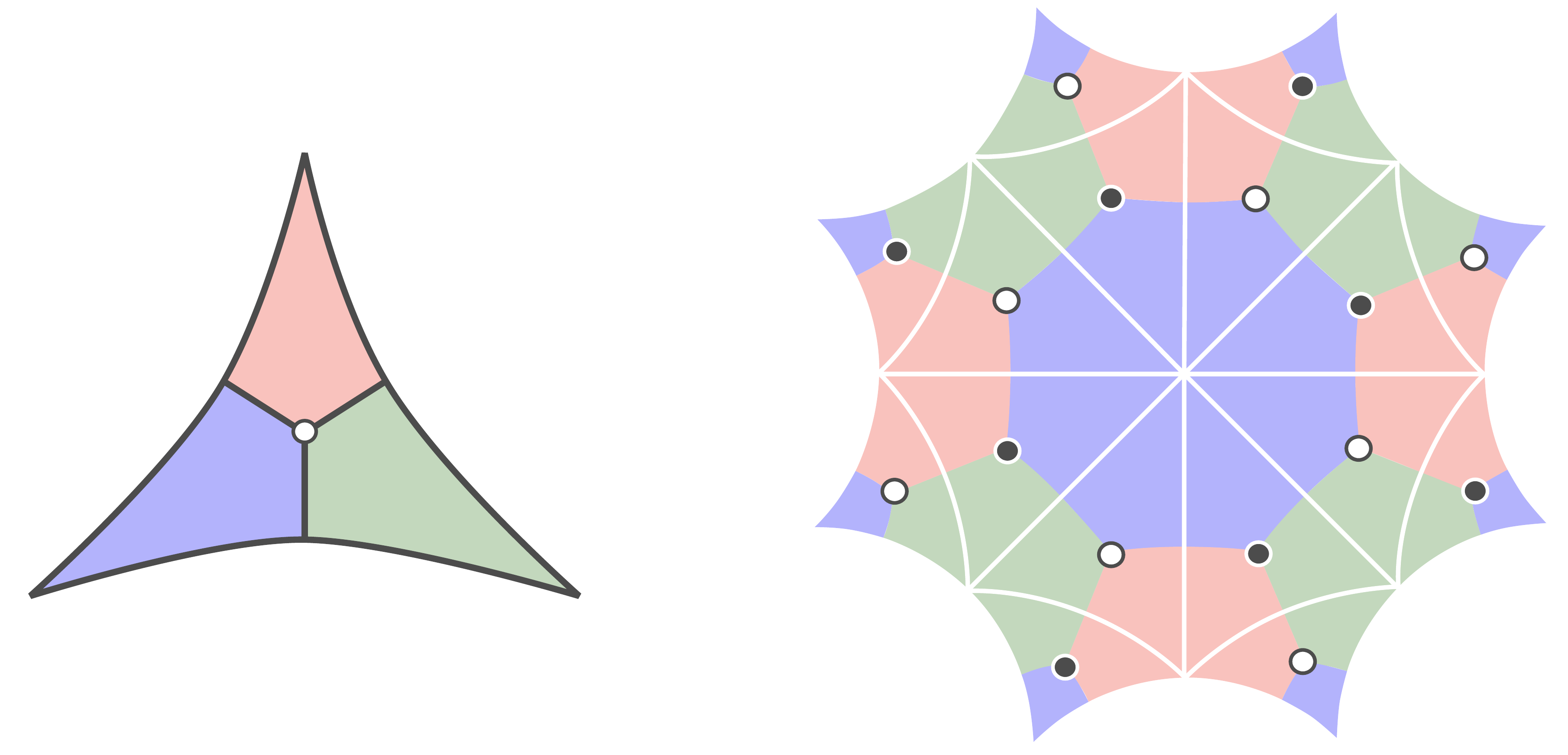}
    \caption{Construction of colour code tilings via the Wythoff construction.
    Left:~Fundamental triangle with internal angles $\pi/4$, $\pi/4$ and $\pi/4$ belonging to the hyperbolic triangle group $\Delta(4,4,4)$. 
    Right:~The fundamental domain of the group $\Gamma_B$ defining the Bolza surface $\hyp/\Gamma_B$.
    In the figure opposite sides are identified, so that $\operatorname{dim} H_1(\hyp/\Gamma_B) = 4$.
    The Bolza surface supports the three-colourable 8.8.8 tiling.
    Neighbouring fundamental triangles of the tiling are related by a reflection along the triangle's side.
    The vertices in the middle of each fundamental triangle swap between black and white with each reflection.
    }
    \label{fig:wythoff}
\end{figure}

As discussed in \Cref{sec:hyperbolic_colour_code_tilings}, we can use a compactification procedure to construct a finite tiling of a closed hyperbolic surface from an infinite tiling~\cite{breuckmann2016constructions,breuckmann2018phd}.
Concretely, for a triangle group $\Delta(p,q,t)$ we can find a normal subgroup $\Gamma \subset \Delta(p,q,t)$ that has no fixed points and gives a finite quotient group $\Delta(p,q,t)/\Gamma$, see~\cite{breuckmann2016constructions,breuckmann2018phd}.
The quotient group $G_{p,q,t}^+\coloneqq\Delta(p,q,t)/\Gamma$ can then be used to define uniform tilings using the Wythoff construction, and has presentation
\begin{multline}
G_{p,q,t}^+=\langle a,b,c~|~a^2=b^2=c^2=(ab)^t=(bc)^q\\=(ca)^p=r_1=\cdots=r_v=e \rangle
\end{multline}
where here the additional relations $r_1,\ldots,r_v$ are the generators of $\Gamma$.

When~$\Gamma \subset \Delta(p,q,t)$ only contains elements consisting of an even number of reflections, the resulting surface is orientable and the graph of the tiling is bipartite.
This is because every vertex is labelled by an element of the reflection group $h\in \Delta(p,q,t) / \Gamma$ and, assuming that $\Gamma$ contains only even parity elements, the assignment of a parity to each $h$ is well-defined.
An example of such a finite tiling (of the Bolza surface) is given in \Cref{fig:wythoff}, where the parity of reflections of vertices are given by the colour black and white.

\subsection{Properties of uniform tilings}

Colour code tilings obtained using the Wythoff construction as described in \Cref{sec:wythoff} are uniform tilings, which can be described by their \textit{vertex configuration}, a sequence of numbers giving the number of sides of the faces around a vertex.
For example, each vertex of a degree-three uniform tiling with vertex configuration $r.g.b$ has three faces around it, with $r$, $g$ and $b$ sides.
When the tiling is constructed using the triangle group $\Delta(p,q,t)$ with a generator point in the middle of the fundamental triangle, we have that $r=2p$, $g=2q$ and $b=2t$.
Uniform tilings of hyperbolic surfaces satisfy $1/r+1/g+1/b<1/2$, whereas a Euclidean tiling (e.g.~6.6.6 or 4.8.8) satisfies $1/r+1/g+1/b=1/2$.
Since we only consider colour code tilings, we can equivalently define $r$, $g$ and $b$ to be the number of sides that red, green and blue faces have in the uniform tiling, respectively.
We construct Floquet codes from $6.6.6$ (honeycomb) tilings, as well as $8.8.8$, $4.8.10$ and $4.10.10$ hyperbolic tilings in this work; see \Cref{fig:honeycomb_and_hyperbolic_floquet_checks} for an illustration of 6.6.6 and 8.8.8 tilings as well as \Cref{fig:wythoff_generator_points} (left) for a 4.8.10 tiling.
Let us denote the set of faces with $r$, $g$ and $b$ sides in an $r.g.b$ uniform tiling as $F_r$, $F_g$ and $F_b$, respectively.
Note that this partition of the faces $F=F_r\cup F_g \cup F_b$ is well defined by the face-colouring even if $r$, $g$ and $b$ are not unique.
We also denote the edge set by $E$ and the vertex set by $V$.
The tiling has $|E|=3|V|/2$ edges, and since $r|F_r|=g|F_g|=b|F_b|=|V|$, we have that 
\begin{equation}
    |F|=|V|\left(\frac{1}{r}+\frac{1}{g}+\frac{1}{b}\right).
    \label{eq:num_faces_trivalent}
\end{equation}
    
If a tiling $\mathcal{T}=(V, E, F)$ has a single connected component and tiles a closed surface, it can be shown that the dimension of the first homology group of the cell complex associated with the tiling has dimension
\begin{equation}\label{eq:num_logical_qubits}
    \dim H_1=2-\chi=2-|V|+|E|-|F|
\end{equation}
 where here $\chi\coloneqq |V|-|E|+|F|$ is the Euler characteristic of the tiling.
 We refer the reader to Ref.~\cite{breuckmann2018phd} for further background on homology theory relevant to quantum codes derived from tilings.
If $\mathcal{T}$ is further a uniform $r.g.b$ colour code tiling we have that
\begin{equation}\label{eq:uniform_dim_H_1}
    \dim H_1=2-|V|+|E|-|F|=|V|\left(\frac{1}{2}-\frac{1}{r}-\frac{1}{g}-\frac{1}{b}\right)+2.
\end{equation}
Note that for an orientable surface with genus $w$, it also holds that $\dim H_1=2w$.
The derived Floquet code will encode $k = \dim H_1$ logical qubits into $n = |V|$ physical qubits.
Hence, \Cref{eq:uniform_dim_H_1} implies that this family of codes has finite rate and that the proportionality depends on the plaquette stabiliser weights $r$, $g$, $b$.

\subsection{Semi-hyperbolic colour code tilings}
\label{sec:semi_hyperbolic_tilings}

\begin{figure}
    \centering
    \includegraphics[width=0.7\linewidth]{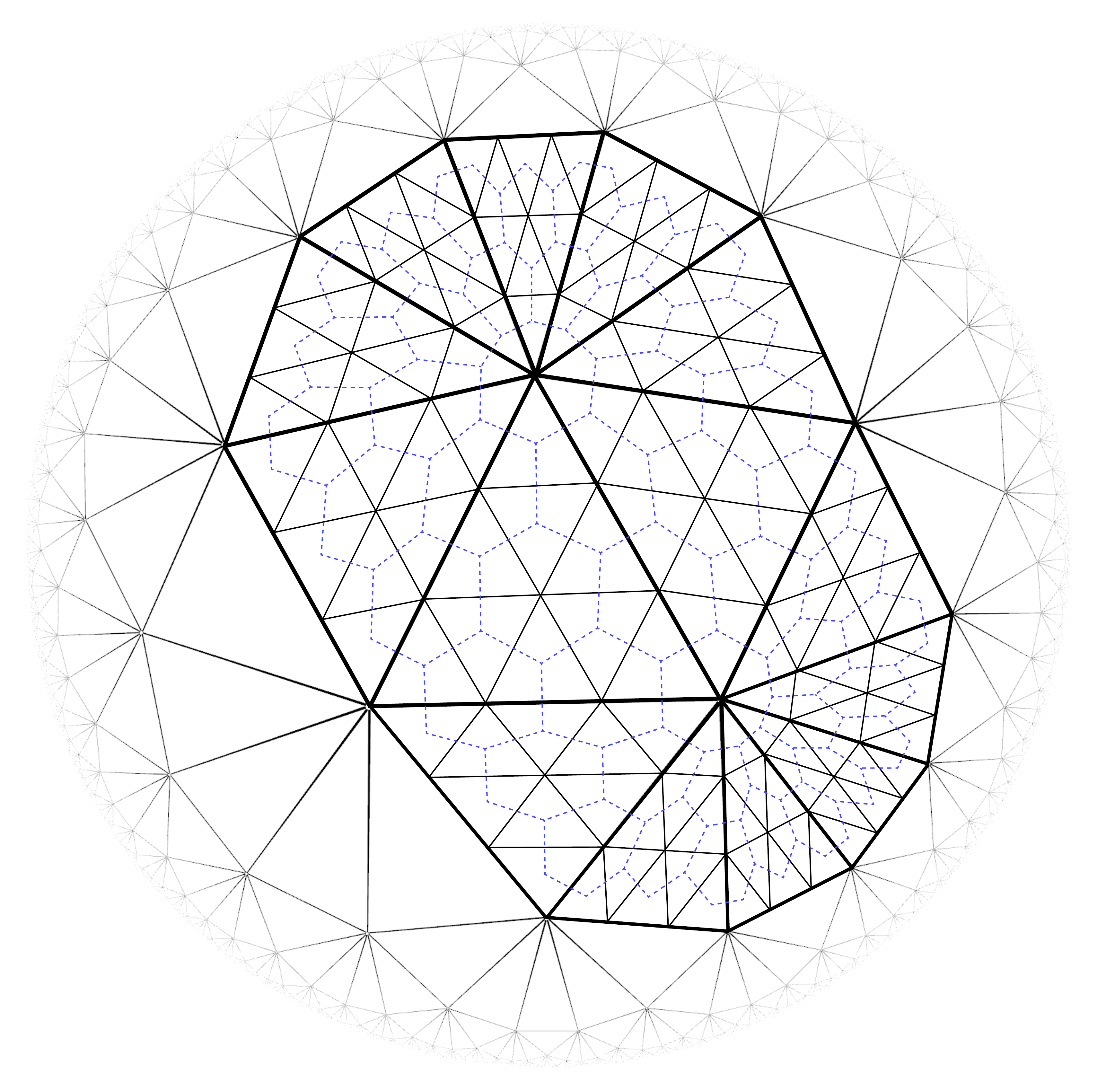}
    \caption{Obtaining a semi-hyperbolic colour code tiling by fine-graining a $3^8$ tiling and then taking its dual. We fine-grain by tiling each face of the $3^8$ tiling with a triangular lattice, such that each side of each face in the $3^8$ tiling becomes subdivided into $l$ edges (shown for a subset of the faces here, with $l=3$). We take the dual of this lattice (blue dashed lines) to obtain a semi-hyperbolic colour code tiling of hexagons and octagons.}
    \label{fig:semi_hyperbolic_tiling}
\end{figure}

We also construct \textit{semi-hyperbolic} colour code tilings, which interpolate between hyperbolic and Euclidean tilings, using a fine-graining procedure.
Our semi-hyperbolic colour code tilings are inspired by the semi-hyperbolic tilings introduced in Ref.~\cite{breuckmann2017semihyperbolic}, which used a slightly different method of fine-graining applied to $4.4.4.4.4$ tilings (instead of colour code tilings) to improve the distance-scaling of standard hyperbolic surface codes.
A similar method was also used to construct subsystem semi-hyperbolic codes in Ref.~\cite{higgott2021subsystem}.
As we will show in \Cref{sec:constructions}, these tilings can be used to obtain Floquet codes with improved distance scaling relative to those derived from purely hyperbolic tilings, while retaining an advantage over honeycomb codes.

A semi-hyperbolic colour code tiling $\mathcal{T}_l=(V_l,E_l,F_l)$ is defined from a \textit{seed} colour code tiling $\mathcal{T}=(V,E,F)$ as well as a parameter $l$, which determines the amount of fine-graining.
To construct $\mathcal{T}_l$ we first take the dual of the seed tiling $\mathcal{T}$, which we recall we denote by $\mathcal{T}^*=(V^*,E^*,F^*)$.
Since $\mathcal{T}$ is a colour code tiling and therefore 3-valent, all the faces of $\mathcal{T}^*$ are triangles, and we have that $|V^*|=|F|$, $|E^*|=|E|$ and $|F^*|=|V|$. 
If $\mathcal{T}$ is an $8.8.8$ (also denoted $8^3$) tiling then $\mathcal{T}^*$ is a $3^8$ uniform tiling, i.e.~with 8 triangles meeting at each vertex.
We construct a new tiling $\mathcal{T}_l^*$ which fine-grains $\mathcal{T}^*$ by tiling each face of $\mathcal{T}^*$ with a triangular lattice, as shown in \Cref{fig:semi_hyperbolic_tiling}, such that each edge in $\mathcal{T}^*$ is subdivided into $l$ edges in $\mathcal{T}_l^*$.
Finally, we take the dual of $\mathcal{T}_l^*$ to obtain our semi-hyperbolic colour code tiling $\mathcal{T}_l$.
The example of this procedure given in \Cref{fig:semi_hyperbolic_tiling} starts from $\mathcal{T}^*$ (here a $3^8$ tiling) using $l=3$ to obtain a colour code tiling of hexagons and octagons.

Given an $r.g.b$ uniform colour code tiling $\mathcal{T}$, the semi-hyperbolic colour code tiling $\mathcal{T}_l$ derived from it using this procedure has a number of vertices, edges and faces given by:
\begin{align}
    |V_l|&=l^2|V|,\label{eq:semi_hyperbolic_num_vertices}\\
    |E_l|&=\frac{3l^2|V|}{2},\label{eq:semi_hyperbolic_num_edges}\\
    |F_l|&=\left(\frac{l^2}{2} - \frac{1}{2} + \frac{1}{r}+\frac{1}{g}+\frac{1}{b}\right)|V|.\label{eq:semi_hyperbolic_num_faces}
\end{align}
Here \Cref{eq:semi_hyperbolic_num_vertices} can be understood by noticing that the semi-hyperbolic construction replaces each face in the dual tiling $\mathcal{T}*$ (i.e.~a vertex in $\mathcal{T}$) with a triangular tiling of $l^2$ faces in $\mathcal{T}_l^*$ (each a vertex in $\mathcal{T}_l$).
\Cref{eq:semi_hyperbolic_num_edges} follows from the fact that $\mathcal{T}_l$ is trivalent and hence $3|V_l|=2|E_l|$.
We derive \Cref{eq:semi_hyperbolic_num_faces} using the dual lattice, an example of which is shown in \Cref{fig:semi_hyperbolic_tiling}.
i.e.~We will find the number number of vertices $|V_l^*|$ in~$\mathcal{T}_l^*$ in terms of the number of number of faces $|F^*|$ in~$\mathcal{T}^*$.
We note that $\mathcal{T}_l^*$ still contains $|F^*|(1/r+1/g+1/b)$ vertices (see \Cref{eq:num_faces_trivalent}) corresponding to the vertices from $\mathcal{T}^*$.
Additionally we have added $l-1$ additional vertices along each of the $|E^*|=3|F^*|/2$ edges of $\mathcal{T}^*$, as well as $(l-1)(l-2)/2$ vertices fully contained within the interior of each of the original faces of $\mathcal{T}^*$.
Combining these contributions and switching back to the primal tiling $\mathcal{T}_l$ we obtain \Cref{eq:semi_hyperbolic_num_faces}.

Since we have not modified the topology of the surface, the dimension of the first homology group $\dim H_1=2-|V_l|+|E_l|-|F_l|$ is independent of $l$ and still given by \Cref{eq:uniform_dim_H_1}.

\section{Floquet codes}\label{sec:floquet_codes}

Floquet codes, introduced by Hastings and Haah~\cite{Hastings2021dynamically}, are quantum error correcting codes that are implemented by measuring two-qubit Pauli operators (they have weight-two checks).
In Ref.~\cite{Hastings2021dynamically} the authors introduced and studied the honeycomb code, which is a Floquet code defined from a hexagonal tiling of a torus, inspired by Kitaev's honeycomb model~\cite{kitaev2006anyons}.
Shortly after, Vuillot showed that Floquet codes can be defined from any colour code tiling, and observed that the use of hyperbolic colour code tilings would lead to Floquet codes with a finite encoding rate and logarithmic distance~\cite{vuillot2021planar}.

We will focus our attention on Floquet codes derived from colour code tilings of \textit{closed} surfaces. 
It is also possible to construct planar Floquet codes; we refer the reader to Refs.~\cite{haah2022boundaries, paetznick2023performance,gidney2022benchmarking, davydova2023quantum,aasen2023measurement} for details.
The definition of Floquet codes we use is consistent with Refs.~\cite{vuillot2021planar, gidney2021fault}, which is related to the original definition of Ref.~\cite{Hastings2021dynamically} by a local Clifford unitary.

\subsection{Checks and stabilisers}\label{sec:checks_and_stabilisers}

We use the colour code tiling $\mathcal{T}$ to define a Floquet code, which we will denote by $\mathcal{F}(\mathcal{T})$.
Each vertex in~$\mathcal{T}$ represents a qubit in $\mathcal{F}(\mathcal{T})$ and each edge in $\mathcal{T}$ represents a two-qubit check operator.
There are three types of check operators: $X$-checks (red edges), $Y$-checks (green edges) and $Z$-checks (blue edges).
Specifically, the check operator for an edge $e=(v_1, v_2)$ is defined to be $P_e\coloneqq P_{v_1}^{C(e)}P_{v_2}^{C(e)}$, where here $P_q^c$ denotes a Pauli operator acting on qubit~$q$ labelled by the colour $c\in\{R,G,B\}$ which determines the Pauli type.
i.e.~$P_q^B$, $P_q^G$ and~$P_q^R$ denote Pauli operators $X_q$, $Y_q$ and $Z_q$, respectively.
For the Floquet codes we consider in this work, the Pauli type of an edge is fixed and is determined by its colour, and we will sometimes refer to checks of a given colour as $c$-checks, for $c\in\{R,G,B\}$.
However, we note that alternative variants of Floquet codes have been introduced for which the Pauli type of an edge measurement is allowed to vary in the schedule (e.g.~Floquet colour codes~\cite{kesselring2022anyon}).

Each stabiliser generator of the Floquet code consists of a cycle of checks surrounding a face.
In other words we associate a stabiliser generator $P_f\coloneqq \prod_{e\subset f}P_e$ with each face~$f$.
Stabiliser generators on red, green and blue faces are $X$-type, $Y$-type and $Z$-type respectively.
We sometimes refer to these stabiliser generators as the \textit{plaquette stabilisers} for clarity.
Note that the plaquette stabilisers commute with all the checks.
See \Cref{fig:honeycomb_and_hyperbolic_floquet_checks} for examples of the checks and stabilizers for Floquet codes defined on hexagonal (6.6.6) and hyperbolic (8.8.8) lattices.

\subsection{The schedule and instantaneous stabiliser group}

The Floquet code is implemented by repeating a \textit{round} of measurements, where each round consists of measuring all red checks, then all green checks and finally all blue checks.
We refer to the measurement of all checks of a given colour as a \textit{sub-round} (so each round contains three sub-rounds).

Once the steady state is reached, after fault-tolerant initialisation of the Floquet code, the state is always in the joint $+1$-eigenspace of the plaquette stabilisers defined in \Cref{sec:checks_and_stabilisers}, regardless of which sub-round was just measured.
This property allows measurements of the plaquette stabilisers to be used to define detectors for decoding (see \Cref{sec:detectors_and_decoding}).
However, the full stabiliser group of the code at a given instant includes additional stabilisers corresponding to the check operators measured in the most-recent sub-round.
We refer to the full stabiliser group of the code after a given sub-round as the \textit{instantaneous stabiliser group} (ISG)~\cite{Hastings2021dynamically}.
In the steady state of the code, just after measuring a sub-round of checks with colour $c\in{R,G,B}$, the ISG is generated by all the plaquette stabilisers, as well as the check operators of colour $c$.
The phase of each check operator in the ISG is given by its measurement outcome in the most-recent sub-round.
We will sometimes refer to the ISG immediately after measuring $c$-checks as $\mathrm{ISG}(c)$.

\subsection{The embedded homological code}

The state of a Floquet code can be mapped by a constant-depth unitary circuit to a 2D homological code, which we refer to as the embedded homological code.
This mapping is shown in Ref.~\cite{Hastings2021dynamically} for the honeycomb code and in Ref.~\cite{vuillot2021planar} for general colour code tilings.
The embedded homological code can be useful to understand and construct the logical operators of the code, as we will explain in \Cref{sec:logicals}.
As an example, consider the state of the Floquet code immediately after a red sub-round, during which any red edge $e=(u, v)$ has participated in an $X_uX_v$ measurement.
This $X_uX_v$ measurement projects the qubits $u$ and $v$ into a two-dimensional subspace, which we can consider an effective qubit in an embedded 2D homological code.
The effective $X$ and $Z$ operators associated with this effective qubit are defined to be $\hat{X}_e\coloneqq Y_uY_v$ (or $Z_uZ_v$) and $\hat{Z}_e\coloneqq X_uI_v$ (or $I_uX_v$), respectively.

\begin{figure}
    \centering
    \includegraphics[width=0.95\linewidth]{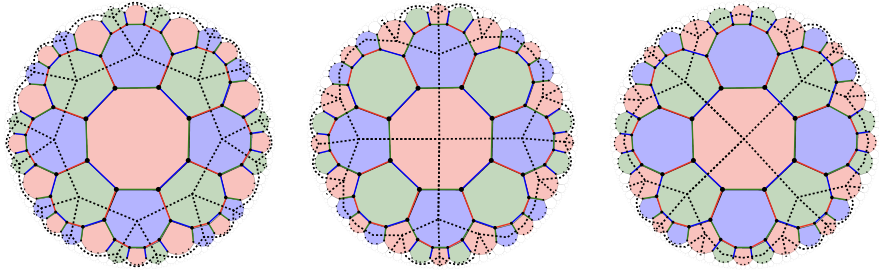}
    \caption{The dashed lines in the left, middle and right diagrams of the 8.8.8 tiling show the red ($\mathcal{T}^*_R$), green ($\mathcal{T}^*_G$) and blue ($\mathcal{T}^*_B$) restricted lattices, respectively.
    The restricted lattice $\mathcal{T}^*_c$ of colour $c\in \{R,G,B\}$ defines the embedded homological code after the $c$-checks are measured.
    Each edge, face and vertex in the restricted lattice corresponds to an effective qubit, plaquette stabiliser or site stabiliser in the embedded homological code, respectively.
    All three restricted lattices are regular tessellations, with four octagons meeting at each vertex.}
    \label{fig:8_3_tessellation_restricted_lattices}
\end{figure}

More concretely, we denote by $\mathcal{T}^*$ the dual of the colour code tiling $\mathcal{T}$, where we define a vertex in $\mathcal{T}^*$ for each face of $\mathcal{T}$, and two nodes in $\mathcal{T}^*$ are connected by an edge iff their corresponding faces in $\mathcal{T}$ share an edge.
Furthermore, the colour of a node in $\mathcal{T}^*$ is given by the colour of the corresponding face in $\mathcal{T}$.
We then define the restricted lattice $\mathcal{T}^*_c$ for $c\in\{R,G,B\}$ to be the subgraph of $\mathcal{T}^*$ where all nodes of colour $c$ (and their adjacent edges) have been removed~\cite{Kubica2023efficientcolorcode}.
We refer to $\mathcal{T}^*_R$ as the red restricted lattice (and similarly for green and blue).
See \Cref{fig:8_3_tessellation_restricted_lattices} for examples of the red, green and blue restricted lattices for an 8.8.8 hyperbolic tiling.

The embedded 2D homological code $\mathcal{C}(\mathcal{T}^*_c)$ associated with the Floquet code after measuring checks of colour $c$ is defined from $\mathcal{T}^*_c$ by associating qubits with the edges, $Z$-checks with the faces and $X$-checks with the vertices.
An effective $\hat{X}$ operator on an effective qubit in $\mathcal{T}^*_c$ corresponds to a physical $P^c\otimes I$ operator on the Floquet code data qubits and similarly an effective $\hat{Z}$ operator corresponds to a physical $P^{\bar{c}}\otimes P^{\bar{c}}$ operator, where here $\bar{c}\in \{R, G, B\}\setminus \{c\}$.
As an example of this mapping, notice that a red $X^{\otimes 8}$ plaquette stabiliser in \Cref{fig:8_3_tessellation_restricted_lattices} corresponds to a $\hat{Z}^{\otimes 8}$ plaquette operator in $\mathcal{C}(\mathcal{T}^*_R)$ and to an $\hat{X}^{\otimes 4}$ site operator in $\mathcal{C}(\mathcal{T}^*_G)$ or $\mathcal{C}(\mathcal{T}^*_B)$.

\subsection{Logical operators}\label{sec:logicals}

\begin{figure}
    \centering
    \includegraphics[width=\linewidth]{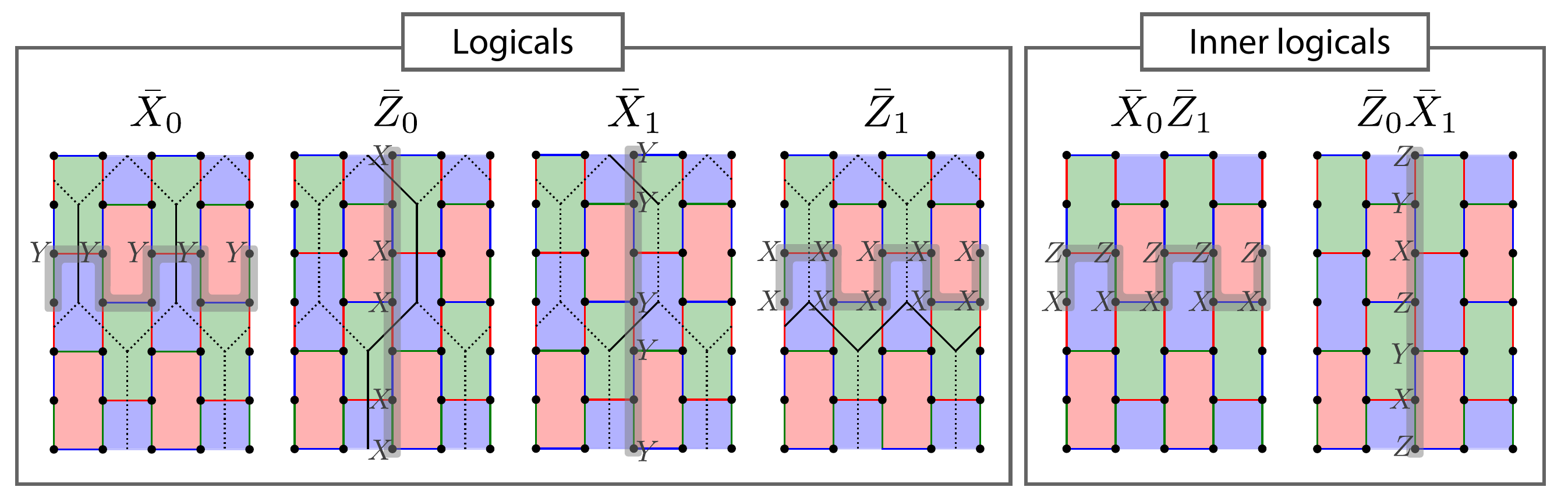}
    \caption{Left: Logical operators of the two logical qubits of the toric honeycomb code after measuring the red checks. These representatives of the logical operators commute with the round of green checks that follows. The overlaid black lattice (with dotted and solid lines) is the tiling of the corresponding embedded toric code, which has an effective qubit associated with each edge. The grey highlighted region is the logical path of the corresponding logical operator. The nontrivial support of the logical operator lies along its logical path in all sub-rounds, even though the Pauli operator itself changes. Right: The inner logical operators, which are products of the logical operators shown on the left. Each inner logical is a homologically non-trivial path of check operators which commutes with all check operators.}
    \label{fig:honeycomb_logicals}
\end{figure}

We can derive a set of logical operators of a Floquet code from the logical operators of the embedded 2D homological code~\cite{Hastings2021dynamically, vuillot2021planar}, as shown in \Cref{fig:honeycomb_logicals} for the toric honeycomb code.
An unusual property of Floquet codes is that the logical operators generally do not commute with all of the checks.
For example, none of the logical operators shown in \Cref{fig:honeycomb_logicals} (left) commute with the blue checks.
Fortunately we can still preserve the logical operators throughout the schedule by updating the logical operators after each sub-round, multiplying an element of the ISG into each of them, such that they commute with the next sub-round of check operator measurements.
In this section we will give a suitable basis for the logical operators and show how they are updated in each sub-round to commute with the checks.
This was explained in Ref.~\cite{gidney2021fault} for the toric honeycomb code (see their Figure~1) whereas our description here is applicable to any Floquet code derived from a colour code tiling of a closed surface (Euclidean or hyperbolic).

We will define a basis for the logical operators of a Floquet code $\mathcal{F}(\mathcal{T})$ from the logical operators of $\mathcal{C}(\mathcal{T}^*_R)$, the embedded homological code immediately after a red sub-round.
After the first red sub-round of $\mathcal{F}(\mathcal{T})$, each $\bar{X}_i$ logical of~$\mathcal{F}(\mathcal{T})$ is defined to be an $\bar{X}_i$ logical of~$\mathcal{C}(\mathcal{T}^*_R)$ (a homologically non-trivial co-cycle), and each $\bar{Z}_i$ logical of $\mathcal{F}(\mathcal{T})$ is defined to be an $\bar{Z}_i$ logical of $\mathcal{C}(\mathcal{T}^*_R)$ (a homologically non-trivial cycle).
See \Cref{fig:honeycomb_logicals} (left) for an example using the toric honeycomb code.
With this choice of logical basis, none of the representatives of the~$\bar{X}_i$ or $\bar{Z}_i$ operators commute with all the check operators, so they are what Ref.~\cite{Hastings2021dynamically} refers to as \textit{outer logical operators}.

We need representatives of logical $\bar{X}_i$ and $\bar{Z}_i$ operators that commute with the green sub-round that immediately follows, however the definition we have just given does not guarantee this property.
This is because our choice of logicals so far is only defined up to an arbitrary element of $\mathrm{ISG}(R)$, which is generated by plaquette stabilisers as well as red checks, and the red checks do not generally commute with the green checks.
We will now show how we can choose a representative for each $\bar{X}_i$ and $\bar{Z}_i$ operator that is guaranteed to commute with the subsequent green checks.

\begin{figure}
    \centering
    \includegraphics[width=\linewidth]{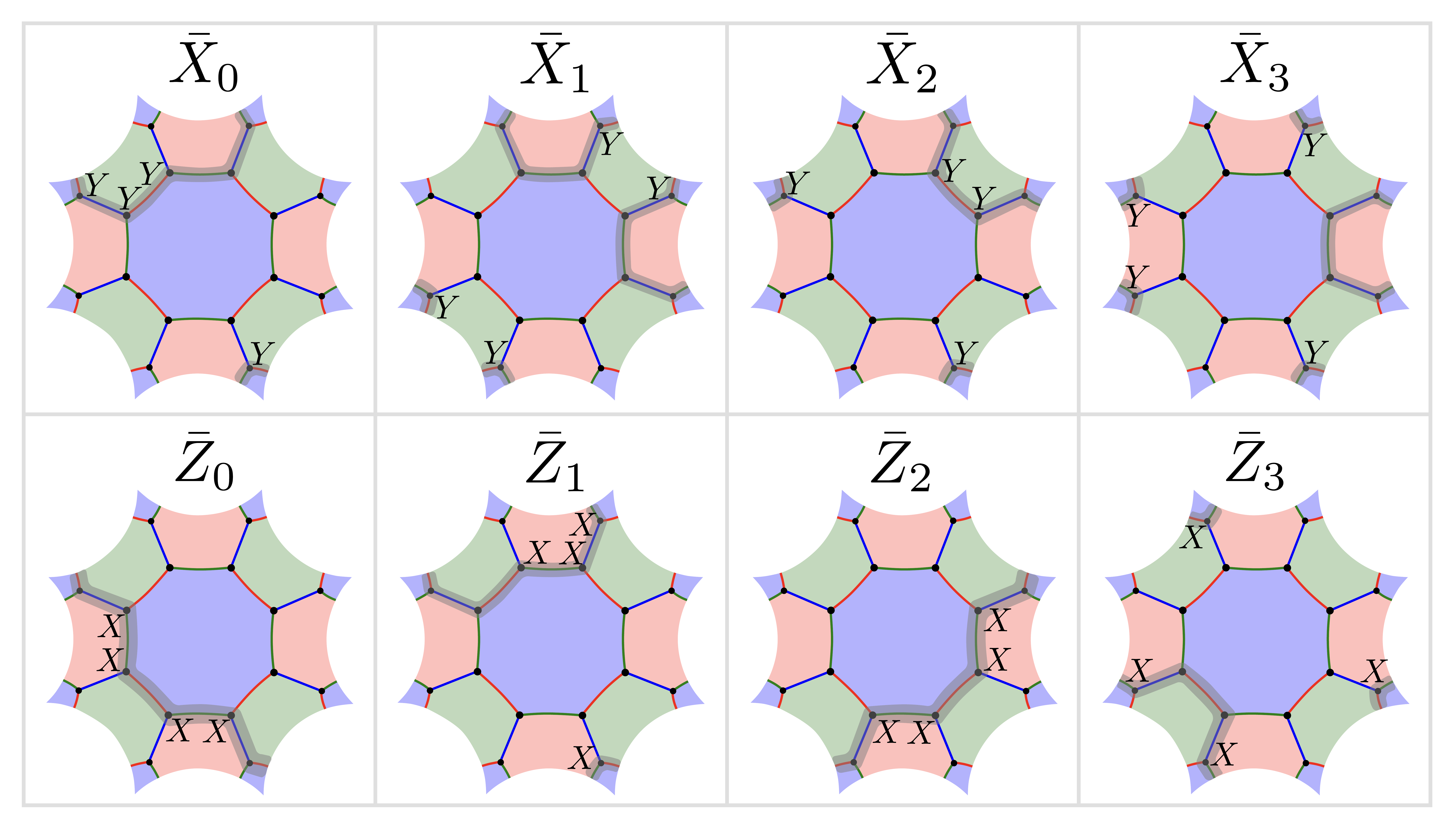}
    \caption{A symplectic basis for the logical operators of the hyperbolic Floquet code derived from the 8.8.8 tiling of the Bolza surface, which has genus 2 and encodes 4 logical qubits into 16 physical data qubits. 
    Opposite sides of the tiling are identified. 
    The logical $\bar{X}$ and $\bar{Z}$ operators of logical qubit $i$ are denoted by $\bar{X}_i$ and $\bar{Z}_i$, respectively. For each logical, the grey highlighted path is its associated homologically non-trivial logical path, which defines how the logical is updated in each sub-round (see \Cref{fig:bolza_logical_over_schedule}).}
    \label{fig:bolza_logicals}
\end{figure}

For each logical $\bar{X}_i$ or $\bar{Z}_i$ operator we associate a \textit{logical path} $P(\bar{X}_i)$ or $P(\bar{Z}_i)$ (highlighted in grey in \Cref{fig:honeycomb_logicals} and \Cref{fig:bolza_logicals}), a homologically non-trivial cycle on $\mathcal{T}$ that includes the nontrivial support of the logical operator.
We can find each logical path using a procedure that \text{lifts} the homologically non-trivial co-cycle (for $\bar{X}_i$) or cycle (for $\bar{Z}_i$) of the logical operator from the restricted lattice to a homologically non-trivial cycle on the colour code tiling $\mathcal{T}$.
We will use the red restricted lattice $\mathcal{T}^*_R$ as an example.
We can lift a \textit{co-cycle} $Q$ in $\mathcal{T}^*_R$ to a logical path $P(X_i)$ in $\mathcal{T}$ by choosing a path that passes only through the red edges that $Q$ crosses and around the borders of red faces.
We can lift a \textit{cycle} $W$ in~$\mathcal{T}^*_R$ to a logical path~$P(Z_i)$ in $\mathcal{T}$ by finding a homologically non-trivial path in $\mathcal{T}$ that passes only through the borders of the blue and green faces in $\mathcal{T}$ associated with the nodes in $W$.
Examples are shown in \Cref{fig:honeycomb_logicals}.

We have now explained how we obtain a logical path (a homologically non-trivial cycle in $\mathcal{T}$) for each logical operator of the Floquet code.
This logical path of a logical operator is fixed throughout the measurement schedule, however the form of the logical Pauli operator does change after the measurement of each sub-round while keeping its support within its logical path.
We now define the form of the logical Pauli operator by defining a mapping $\mathcal{V}_r^B$ which, for a given sub-round $r\in\{0,\ldots,5\}$ and logical basis $B\in\{X,Z\}$, maps a cycle (or set of cycles) in $\mathcal{T}$ to a physical Pauli operator.
We define $\mathcal{V}_r^B$ using the following table:
\begin{center}
\begin{tabular}{ c|c|c } 
 $r\pmod 6$ & $\mathcal{V}_r^X$ & $\mathcal{V}_r^Z$ \\
 \hline
 0 & $Y$ on red edges & $X$ on green edges \\ 
 1 & $Y$ on blue edges & $Z$ on green edges  \\ 
 2 & $X$ on blue edges & $Z$ on red edges \\ 
 3 & $X$ on green edges & $Y$ on red edges \\ 
 4 & $Z$ on green edges & $Y$ on blue edges \\ 
 5 & $Z$ on red edges & $X$ on blue edges \\ 
\end{tabular}
\end{center}
Note that we have $\mathcal{V}_i^X=\mathcal{V}_{i+3\pmod 6}^Z$ and $\mathcal{V}_i^Z=\mathcal{V}_{i+3\pmod 6}^X$.
As an example, given some cycle (or set of cycles) $P$ on the tiling $\mathcal{T}$, the mapping $\mathcal{V}_0^X$ assigns a Pauli $Y$ operator to each qubit at the endpoint of a red edge within $P$ and assigns an identity operaetor to all other qubits.

We then use this mapping to define the logical operator from its logical path. Concretely, we choose the representative of logical $\bar{X}_i$ to be $\mathcal{V}_r^X(P(\bar{X}_i))$ and similarly of logical $\bar{Z}_i$ to be $\mathcal{V}_r^Z(P(\bar{Z}_i))$.
It is straightforward to verify that these are valid representatives of the logical operators and that they always commute with the check operator measurements in the sub-round that immediately follows.

Furthermore, for any face $f$ of $\mathcal{T}$, then both $\mathcal{V}_r^X(f)$ and $\mathcal{V}_r^Z(f)$ are each either the identity or an element of the ISG in sub-round $r$.
Hence, for each of the logical paths $P(\bar{X}_i)$ and $P(\bar{Z}_i)$ we could have picked \textit{any} homologically non-trivial cycle belonging to the same homology class.
This is because any two such choices belonging to the same homology class would differ by a set of faces in $\mathcal{T}$, and hence the corresponding Pauli operators obtained using $\mathcal{V}_r^X$ or $\mathcal{V}_r^X$ would differ by an element of the ISG.
The use of the restricted lattice is helpful to understand the logical operators and paths, and to lift logical operators from the embedded homological code to the Floquet code (as we did for the numerical simulations in this work), but is not strictly necessary for constructing them; we could instead find homologically non-trivial cycles on the colour code lattice directly.
Logical operators and paths for the toric honeycomb code are shown on the left of \Cref{fig:honeycomb_logicals}.
Similarly, logical operators and paths of the hyperbolic Floquet code derived from the Bolza surface (the Bolza Floquet code) are shown in \Cref{fig:bolza_logicals}.

\begin{figure}[t]
    \centering
    \includegraphics[width=0.95\linewidth]{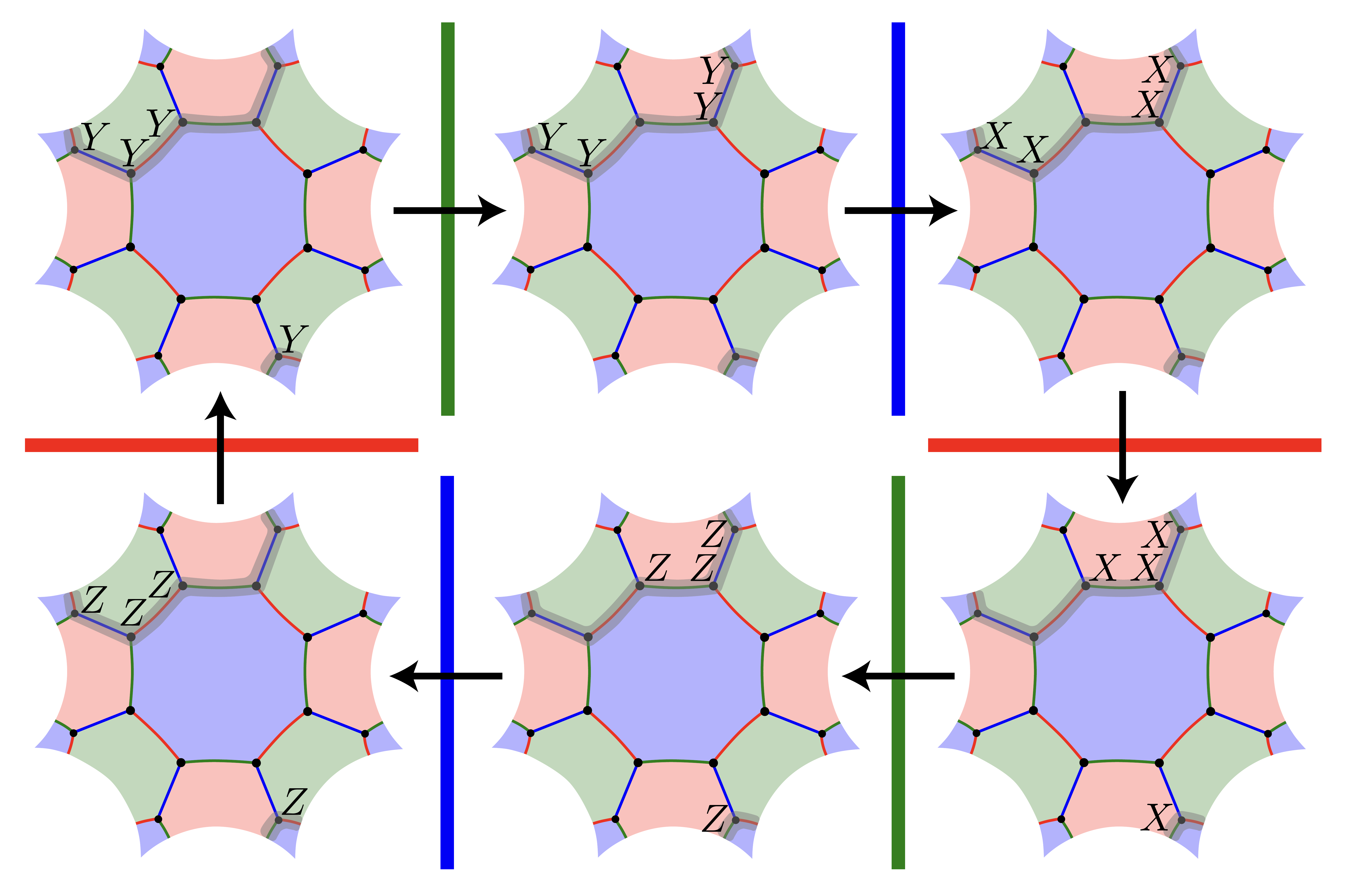}
    \caption{Each logical operator changes over a periodic sequence of six sub-rounds (two full rounds), as shown here for a logical operator ($\bar{X}_0$ in \Cref{fig:bolza_logicals}) of the Bolza Floquet code.
    Opposite sides of the tiling are identified.
    Crossing a bar of colour $c\in\{R,G,B\}$ means measuring the $c$-checks and then multiplying into each logical operator the $c$-checks that lie within its logical path. Note that the $\bar{X}_0$ logical operator (top left) is mapped to the $\bar{Z}_1$ logical operator (bottom right) every three sub-rounds and vice versa. This transition is an automorphism of $\mathrm{ISG}(R)$ which acts on each logical operator by multiplying by the inner logical operator that lies along the same logical path.}
    \label{fig:bolza_logical_over_schedule}
\end{figure}

We have so far only described the form of the logical operators, however we will now explain how the logical operators are updated after each sub-round.
After measuring the checks of colour $c\in\{R,G,B\}$, we multiply into each logical operator the $c$-coloured checks that lie within its logical path.
This maps a logical $\mathcal{V}_r^{X}(P)$ or $\mathcal{V}_r^{Z}(P)$ to $\mathcal{V}_{r + 1\pmod 6}^{X}(P)$ or $\mathcal{V}_{r + 1\pmod 6}^{Z}(P)$, respectively and guarantees that it will commute with the sub-round of checks that immediately follows.
The following table describes the form of the logical operators over the period of 6 sub-rounds:
\begin{center}
\resizebox{\linewidth}{!}{
\begin{tabular}{ c|c|c|c|c|c } 
 $r\pmod 6$ & ISG & Form of $\bar{X}_i$ & $\bar{X}_i$ in $\mathcal{T}^*_c$ & Form of $\bar{Z}_i$ & $\bar{Z}_i$ in $\mathcal{T}^*_c$ \\
 \hline
 0 & $R$ & $\mathcal{V}_0^X(P(\bar{X}_i))$ & co-cycle in $\mathcal{T}^*_R$ & $\mathcal{V}_0^Z(P(\bar{Z}_i))$ & cycle in $\mathcal{T}^*_R$ \\ 
 1 & $G$ & $\mathcal{V}_1^X(P(\bar{X}_i))$ & cycle in $\mathcal{T}^*_G$ & $\mathcal{V}_1^Z(P(\bar{Z}_i))$ & co-cycle in $\mathcal{T}^*_G$ \\ 
 2 & $B$ & $\mathcal{V}_2^X(P(\bar{X}_i))$ & co-cycle in $\mathcal{T}^*_B$ & $\mathcal{V}_2^Z(P(\bar{Z}_i))$ & cycle in $\mathcal{T}^*_B$ \\ 
 3 & $R$ & $\mathcal{V}_3^X(P(\bar{X}_i))$ & cycle in $\mathcal{T}^*_R$ & $\mathcal{V}_3^Z(P(\bar{Z}_i))$ & co-cycle in $\mathcal{T}^*_R$ \\ 
 4 & $G$ & $\mathcal{V}_4^X(P(\bar{X}_i))$ & co-cycle in $\mathcal{T}^*_G$ & $\mathcal{V}_4^Z(P(\bar{Z}_i))$ & cycle in $\mathcal{T}^*_G$ \\ 
 5 & $B$ & $\mathcal{V}_5^X(P(\bar{X}_i))$ & cycle in $\mathcal{T}^*_B$ & $\mathcal{V}_5^Z(P(\bar{Z}_i))$ & co-cycle in $\mathcal{T}^*_B$ \\ 
\end{tabular}}
\end{center}
where here each row describes the form of the logical operators on $\mathcal{T}$ and $\mathcal{T}^*_c$ after a given sub-round $r$, which depends on $r\pmod 6$.
See \Cref{fig:bolza_logical_over_schedule} for the cycle of a logical operator in the Bolza Floquet code, as well as Figure~1 of Ref.~\cite{gidney2021fault} for a similar figure using the honeycomb code.
An $\bar{X}$-type logical operator in $\mathrm{ISG}(R)$ is mapped to a $\bar{Z}$-type logical operator in $\mathrm{ISG}(R)$ (and vice versa) every three sub-rounds by an automorphism of $\mathrm{ISG}(R)$.
The same applies to $\mathrm{ISG}(G)$ and $\mathrm{ISG}(B)$.
Since we must multiply checks into the logicals in each sub-round, the measurement of a logical operator in a Floquet code includes the product of measurement outcomes spanning a sheet in space-time, rather than just a string of measurements in the final round as done for transversal measurement in the surface code.

So far we have only consider outer logical operators, which move after each sub-round.
However there are some logical operators, called \textit{inner logical operators} in Ref.~\cite{Hastings2021dynamically}, which commute with all the checks.
These inner logical operators act non-trivially on the encoded logical qubits of the Floquet code and are products of checks lying along homologically nontrivial cycles of the lattice.
Each inner logical operator is formed from the product of an $\bar{X}$ and a $\bar{Z}$ logical operator associated with the same homologically nontrivial cycle of the lattice.
For example, \Cref{fig:honeycomb_logicals} (right) shows the two inner logical operators, $\bar{X}_0\bar{Z}_1$ and $\bar{Z}_0\bar{X}_1$, in the toric honeycomb code.
One of the four inner logical operator of the Bolza Floquet code can be formed from the product of the $\bar{X}_0$ and $\bar{Z}_1$ logical operators in \Cref{fig:bolza_logicals}.
However, even though the inner logical operators are formed from a product of check operators, the measurement schedule of the Floquet code is chosen such that it never measures the inner logical operators, only the plaquette stabilisers~\cite{Hastings2021dynamically, vuillot2021planar}.

\subsection{Circuits and noise models}

In this work we consider two different noise models, corresponding to the EM3 and SD6 noise models used in Ref.~\cite{gidney2022benchmarking}.
Both noise models are characterised by a noise strength parameter $p\in[0,1]$ (the physical error rate).

\subsubsection{EM3 noise model}

The EM3 noise model (entangling measurement 3-step cycle) assumes that two-qubit Pauli measurements are available natively in the platform (and hence no ancilla qubits are needed), as is the case for Majorana-based architectures~\cite{Chao2020optimizationof,paetznick2023performance}.
When implementing the measurement of a two-qubit Pauli operator $P^c\otimes P^c$, with probability $p$ we insert an error chosen uniformly at random from the set $\{I,X,Y,Z\}^{\otimes 2}\times \{\text{flip}, \text{no flip}\}$.
Here the two qubit Pauli error in $\{I,X,Y,Z\}^{\otimes 2}$ is applied immediately before the measurement, and the ``flip'' operation flips the outcome of the measurement.
We assume each measurement takes a single time step, and hence each cycle of three sub-rounds takes three time steps.
Initialisation of a qubit in the $Z$ basis is followed by an $X$ error, inserted with probability $p/2$.
Similarly, the measurement of a data qubit in the $Z$ basis is preceded by an $X$ error, inserted with probability $p/2$.
Single-qubit initialisation and measurement in the $X$ or $Y$ basis is achieved using noisy $Z$-basis initialisation or measurement and noiseless single-qubit Clifford gates.
We do not need to define idling errors because data qubits are never idle in this noise model.

\subsubsection{SD6 noise model}

The SD6 noise model (standard depolarizing 6-step cycle) facilitates the check measurements via an ancilla qubit for each edge of the colour code tiling.
Therefore, this noise model introduces $|E|=3|V|/2$ ancilla qubits and requires $2.5\times$ more qubits than the $EM3$ noise model for a given colour code tiling $\mathcal{T}$.
Note that the distance in this noise model (the minimum number of independent error mechanisms required to flip a logical observable but no detectors) can be higher than $EM3$ for a given tiling, so we do not necessarily require $2.5\times$ more qubits for a given distance.

Each measurement of a two-qubit check operator $P^c\otimes P^c$ is implemented using CNOT gates and single-qubit Clifford gates, as well as measurement and reset of the ancilla qubit associated with the edge.
We use the same circuit as in Ref.~\cite{gidney2021fault}.
Each CNOT gate is followed by a two-qubit depolarising channel, which with probability $p$ inserts an error chosen uniformly at random from the set $\{I,X,Y,Z\}^{\otimes 2}\setminus \{I\otimes I\}$.
Each single-qubit Clifford gates is followed by a single-qubit depolarising channel, which with probability $p$ inserts an error chosen uniformly at random from the set $\{X,Y,Z\}$.
Single-qubit initialisation is followed by an $X$ error with probability $p$ and single-qubit measurement is preceded by an $X$ error with probability $p$.
Each cycle of three sub-rounds takes six time steps, with CNOT gates, single-qubit gates, initialisation and measurement all taking one time step each.
In each time step in the bulk, each data qubit is either involved in a CNOT gate or a $C_{ZYX}\coloneqq HS$ gate (which maps $Z\rightarrow Y\rightarrow X\rightarrow Z$ under conjugation) and therefore never idles.

\subsection{Implementing hyperbolic and semi-hyperbolic connectivity}

Although hyperbolic and semi-hyperbolic Floquet codes cannot be implemented using geometrically local connections in a planar Euclidean layout of qubits, they can instead be implemented using \textit{biplanar} or \textit{modular} architectures.

In a biplanar architecture, the connections (couplers) between qubits can be partitioned into two layers, such that the couplers within each layer do not cross.
More concretely, consider the qubit connectivity graph $G_c=(V_c,E_c)$, which by definition contains an edge $(q_i,q_j)\in E_c$ if and only if qubits $q_i$ and $q_j$ directly interacted in the quantum circuit implementing the code.
Recall that the \textit{thickness} of a graph is the minimum number of planar graphs that the graph's edge set can be partitioned into.
A graph is biplanar if it has thickness 2; i.e.~$G_c$ is biplanar if we can partition the edge set as $E_c=E_c^1\cup E_c^2$, where the edge sets $E_c^1$ and $E_c^2$ each define a planar graph.
It is known that the thickness of a graph with maximum degree $d$ has thickness at most $t=\lceil \frac{d}{2}\rceil$ (see Corollary 5 of Ref.~\cite{halton1991thickness}, as well as Proposition 1 of Ref.~\cite{tremblay2022constant} for this application to QEC).
Therefore, since $G_c$ has maximum degree 3 for any (semi-)hyperbolic Floquet code, it must also be biplanar.
Note that the subsystem hyperbolic codes of Ref.~\cite{higgott2021subsystem} are also biplanar, since their qubit connectivity graphs always have degree 4.
Biplanarity of the connectivity graph implies that we can fix the position of the qubits and have two layers of connections (e.g.~above and below), each of which is planar.
That both $E_c^1$ and $E_c^2$ can be planar with fixed qubit positions can be understood from Theorem 8 of Ref.~\cite{halton1991thickness}, which shows that a planar graph always has a planar representation with nodes placed in arbitrary positions.

\begin{figure}
    \centering
    \includegraphics[width=0.6\linewidth]{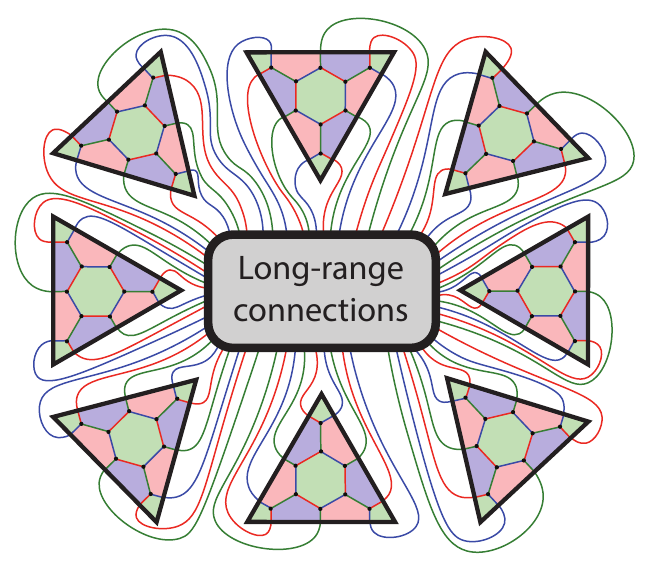}
    \caption{
    A modular architecture for a semi-hyperbolic Floquet code. Each module has the connectivity of a planar Euclidean chip, and arbitrary connectivity is permitted between modules. In this example, each module is the result of fine-graining the neighbourhood of one vertex of the seed tiling $\mathcal{T}$ used to construct an $l=3$ semi-hyperbolic Floquet code $\mathcal{F}(\mathcal{T}_3)$. 
    Each long-range connection corresponds to a two-qubit measurement on qubits in two different modules. These inter-module connections may in general be long-range in order to facilitate the necessary connectivity of the semi-hyperbolic Floquet code.}
    \label{fig:modular}
\end{figure}

We can also implement (semi-)hyperbolic Floquet codes using a modular architecture~\cite{monroe2014large}, in which long-range links connect small modules of qubits.
Examples of qubit platforms that could support modular architectures include ions~\cite{monroe2014large,stephenson2020high,pino2021demonstration}, atoms~\cite{reiserer2015cavity, bluvstein2022quantum} or superconducting qubits~\cite{sahu2022quantum,tu2022high,delaney2022superconducting,Zhu2020waveguide,Imany2022quantum}.
We break up the Floquet code into small modules, where each module supports planar Euclidean connectivity and long-range two-qubit Pauli measurements are used to connect different modules.
For a hyperbolic Floquet code, variable length connections can be used to embed a small region of the lattice into each module~\cite{kollar2019hyperbolic}.
For a semi-hyperbolic Floquet code, local regions of the code have identical connectivity to a honeycomb code; hence, each module can consist of a small hexagonal lattice of qubits obtained from fine-graining the neighbourhood of one vertex of the seed tiling $\mathcal{T}$ that the semi-hyperbolic Floquet code $\mathcal{F}(\mathcal{T}_l)$ was derived from (see \Cref{fig:modular}).
By connecting these modules using long-range two-qubit measurements, we can implement the connectivity required for the tiling of the closed hyperbolic surface.
We note that additional connectivity might be necessary to implement logical gates and move logical qubits into and out of storage (e.g.~using lattice surgery~\cite{breuckmann2017semihyperbolic}), but we leave this research problem to future work.

\subsection{Detectors and decoding}\label{sec:detectors_and_decoding}

In order to decode Floquet codes, we must make an appropriate choice of \textit{detectors}.
Following the definition in the Stim~\cite{Gidney2021stimfaststabilizer} documentation, a detector is a parity of measurement outcomes in the circuit that is deterministic in the absence of errors.
In the literature, detectors have also been referred to as error-sensitive events~\cite{sundaresan2023matching} or checks~\cite{bombin2023logical,delfosse2023spacetime}.
In a Floquet code, each detector in the bulk of the schedule is formed by taking the parity of two consecutive measurements of a plaquette stabiliser.
For example, in the honeycomb code, each plaquette stabiliser is the product of the six edge check measurements around a face, and so a detector is the parity of 12 measurement outcomes.
For a Floquet code derived from an $r.g.b$ tiling, detectors are the parity of $2r$, $2g$ or $2b$ measurement outcomes.
With this choice of detectors, it is known that any Floquet code can be decoded with minimum-weight perfect matching (MWPM)~\cite{dennis2002topological,fowler2014minimum,higgott2023sparse,wu2023fusion} or Union-Find (UF)~\cite{Delfosse2021almostlineartime,huang2020fault} decoders~\cite{Hastings2021dynamically,vuillot2021planar,gidney2021fault} by defining a \textit{detector graph} for a given noise model.
Each node in the detector graph represents a detector and each edge represents an error mechanism that flips the detectors at its endpoints with probability $p$, and has an edge weight given by $\log((1-p)/p)$.
Given this detector graph and an observed syndrome, MWPM or UF can be used to find a minimum-weight or low-weight set of edges consistent with the syndrome, respectively.
Some additional accuracy can be achieved by also exploiting knowledge of hyperedge error mechanisms using a correlated matching~\cite{fowler2013optimal} or belief-matching~\cite{higgott2023improved} decoder, however in this work we use the PyMatching implementation of MWPM~\cite{higgott2023sparse} to decode and use Stim to construct the detector graphs and simulate the circuits~\cite{Gidney2021stimfaststabilizer}.

\section{Constructions}\label{sec:constructions}

In this section we present a performance analysis of our constructions of hyperbolic and semi-hyperbolic codes, including a comparison with honeycomb codes and surface codes.
\Cref{sec:code_parameters} contains our analysis of the code parameters, whereas are simulations using EM3 and SD6 noise models are presented in \Cref{sec:simulations}.

\subsection{Code parameters}\label{sec:code_parameters}

The number of data qubits $n=|V|$ in a Floquet code is determined by the number of vertices in the colour code tiling $\mathcal{T}=(V,E,F)$ and the number of logical qubits is given by the dimension of its first homology group $k=\dim H_1$.
From \Cref{eq:uniform_dim_H_1} and \Cref{eq:semi_hyperbolic_num_vertices} we see that the encoding rate $k/n$ is determined by the plaquette stabiliser weights and the fine-graining parameter $l$.
No ancilla qubits are required for the EM3 noise model ($n_{anc}=0$), whereas for the SD6 noise model we have $n_{anc}=|E|=3n/2$ ancillas.
We denote the total number of physical qubits by $n_{tot}\coloneqq n + n_{anc}$.

The distance of a circuit implementing a Floquet code is the minimum number of error mechanisms required to flip at least one logical observable measurement outcome without flipping the outcome of any detectors.
Therefore the distance depends not only on the colour code tiling used to define the Floquet code, but also the choice of circuit and noise model.
We consider three different types of distance of a Floquet code.
One of these is a ``circuit agnostic'' distance which we will refer to as the Floquet code's \textit{embedded distance}.
The embedded distance $d$ of a Floquet code is defined to be the smallest distance of any of its three embedded 2D homological codes (the smallest homologically non-trivial cycle or co-cycle in $\mathcal{T}^*_R$, $\mathcal{T}^*_G$ or $\mathcal{T}^*_B$).
The embedded distance is more efficient to compute than the distance of a Floquet code circuit, since it involves shortest path searches on the 2D restricted lattices (using the method described in Appendix B of ~\cite{breuckmann2017semihyperbolic}), rather than on the much larger 3D detector graph associated with a specific circuit over many rounds (which we compute using \verb|stim.Circuit.shortest_graphlike_error| in Stim~\cite{Gidney2021stimfaststabilizer}).
We also consider the EM3 distance $d_e$ of a Floquet code, which is the graphlike distance of its circuit for an EM3 noise model.
Empirically, we find that the EM3 distance of a Floquet code almost always matches its embedded distance (see \Cref{tab:hyperbolic_floquet_parameters}), even though the embedded distance does not consider measurement errors or the details of a circuit-level noise model.
Finally, we define the SD6 distance $d_s$ to be the graphlike distance of the Floquet codes circuit for an SD6 noise model.
While the SD6 noise model requires additional qubits (an ancilla qubit for each two-qubit check operator measurement), we note that the SD6 distance can in general be higher than the EM3 distance for a given Floquet code.
This can be explained by the fact, for the EM3 noise model, a single error mechanism can result in any of the 15 nontrivial two-qubit Pauli operators acting on a pair of data qubits in the support of a pair measurement. 
In contrast, for the SD6 noise model, only a subset of these two-qubit Pauli operators can result from a single error mechanism.

To achieve an EM3 distance $d_e$ for either the toric or planar honeycomb code, we need at least $2d_e$ columns of qubits and $3d_e$ rows, and hence use a patch with dimensions $2d_e\times 3d_e$ using $n_{tot}=n=6d_e^2$ physical qubits~\cite{gidney2021fault,haah2022boundaries,gidney2022benchmarking,paetznick2023performance}.
See \Cref{fig:honeycomb_and_hyperbolic_floquet_checks} for an example of a $4\times 6$ toric honeycomb code ($d_e=2$).

\subsubsection{Hyperbolic Floquet codes}

\begin{table}
\caption{Parameters of some of the hyperbolic Floquet codes we have constructed. Here $n$, $k$ and $d$ are the number of physical data qubits, number of logical qubits and embedded distance, respectively. The EM3 distance $d_e$ and SD6 distance $d_s$ are the graphlike distances of 16-round circuits using EM3 and SD6 noise models respectively, computed using Stim~\cite{Gidney2021stimfaststabilizer}. We omit $d_s$ where the calculation is too computationally expensive. We only include a code if $d$, $d_e$ or $d_s$ is at least as large as the corresponding distance for all smaller codes in the family.}
\label{tab:hyperbolic_floquet_parameters}
\centering
    \begin{minipage}[t]{.4\linewidth}
\centering
\begin{tabular}[t]{|Sc|Sc|Sc|Sc|Sc|Sc|}
\multicolumn{6}{c}{8.8.8}\\
\hline
$n$ & $k$ & $d$ & $kd^2/n$ & $d_{e}$ & $d_{s}$ \\
\hline
    16 &    4 &  2 & 1.00 &  2 &  2 \\
    32 &    6 &  2 & 0.75 &  2 &  3 \\
    64 &   10 &  2 & 0.62 &  2 &  4 \\
   256 &   34 &  4 & 2.12 &  4 &  4 \\
   336 &   44 &  3 & 1.18 &  3 &  6 \\
   336 &   44 &  4 & 2.10 &  4 &  6 \\
   512 &   66 &  4 & 2.06 &  4 &  6 \\
   720 &   92 &  4 & 2.04 &  4 &  6 \\
 1,024 &  130 &  4 & 2.03 &  4 &  6 \\
 1,296 &  164 &  4 & 2.02 &  4 &  6 \\
 1,344 &  170 &  6 & 4.55 &  - &  6 \\
 2,688 &  338 &  6 & 4.53 &  - &  - \\
 4,896 &  614 &  6 & 4.51 &  - &  - \\
 5,376 &  674 &  6 & 4.51 &  - &  - \\
 5,760 &  722 &  6 & 4.51 &  - &  - \\
\hline
\end{tabular}
\end{minipage}
\hspace{0.01\linewidth}
\begin{minipage}[t]{.4\linewidth}
\centering
\begin{tabular}[t]{|Sc|Sc|Sc|Sc|Sc|Sc|}
\multicolumn{6}{c}{4.10.10}\\
\hline
$n$ & $k$ & $d$ & $kd^2/n$ & $d_{e}$ & $d_{s}$ \\
\hline
   120 &    8 &  3 & 0.60 &  3 &  4 \\
   160 &   10 &  4 & 1.00 &  4 &  4 \\
   320 &   18 &  5 & 1.41 &  4 &  6 \\
   600 &   32 &  6 & 1.92 &  6 &  6 \\
 3,600 &  182 &  8 & 3.24 &  - &  - \\
10,240 &  514 &  9 & 4.07 &  - &  - \\
19,200 &  962 & 10 & 5.01 &  - &  - \\
\hline
\end{tabular}

\vspace{0.5cm}

\centering
\begin{tabular}[t]{|Sc|Sc|Sc|Sc|Sc|Sc|}
\multicolumn{6}{c}{4.8.10}\\
\hline
$n$ & $k$ & $d$ & $kd^2/n$ & $d_{e}$ & $d_{s}$ \\
\hline
   240 &    8 &  4 & 0.53 &  4 &  6 \\
   640 &   18 &  6 & 1.01 &  6 &  8 \\
 1,440 &   38 &  8 & 1.69 &  8 &  8 \\
 7,200 &  182 & 10 & 2.53 &  - &  - \\
\hline
\end{tabular}
\end{minipage}
\end{table}

Although planar and toric honeycomb codes can only encode one or two logical qubits, respectively, hyperbolic Floquet codes can encode a number of logical qubits $k$ proportional to the number of physical qubits $n$.
From \Cref{eq:uniform_dim_H_1} we see that a hyperbolic Floquet code has a finite encoding rate $k/n=\dim H_1/|V|>R$, where $R$ is $1/8$, $1/40$ and $1/20$ for the $8.8.8$, $4.8.10$ and $4.10.10$ colour code tilings, respectively.

Hyperbolic surface codes are known to have $O(\log{n})$  distance scaling~\cite{siran2001triangle,fetaya2012bounding,delfosse2013tradeoffs,breuckmann2016constructions} and it follows that the embedded distance of a hyperbolic Floquet code therefore also scales as $d=O(\log{n})$~\cite{vuillot2021planar}.
This leads to the parameters of hyperbolic Floquet codes satisfying $kd^2/n=O(\log^2(n))$, which is an asymptotic improvement over the parameters of Euclidean surface codes and Floquet codes, for which $kd^2/n=O(1)$.

We give the parameters of some of the hyperbolic Floquet codes we constructed in \Cref{tab:hyperbolic_floquet_parameters}.
These belong to three families of hyperbolic Floquet codes, derived from 8.8.8, 4.10.10 and 4.8.10 tessellations.
We have picked these tilings as they have the lowest stabiliser weight and give the highest relative distance, although they give the lowest encoding rate, see discussion in Ref.~\cite{breuckmann2016constructions}.
All of the codes constructed have $kd^2/n$ exceeding that of planar and toric honeycomb codes for which $kd^2/n$ is $1/6$ and $1/3$, respectively.
The largest improvement in parameters is given by the [[19200,962,10]] 4.10.10 hyperbolic Floquet code which has $kd^2/n=5.01$, a $30.1\times$ larger ratio than is obtained using the planar honeycomb code.

\subsubsection{Semi-hyperbolic Floquet codes}\label{sec:semi_hyperbolic_floquet}

One potential concern is that hyperbolic Floquet codes can only achieve error suppression that is polynomial in the system size, owing to their $O(\log{n})$ distance scaling.
However, in this section we will show that we can achieve exponential error suppression using the semi-hyperbolic colour code tilings we introduced in \Cref{sec:semi_hyperbolic_floquet}, which fine-grain the hyperbolic lattice and have $O(\sqrt{n})$ distance scaling, while still retaining an advantage over Euclidean Floquet codes.

\begin{figure}
    \centering
    \includegraphics[width=0.99\linewidth]{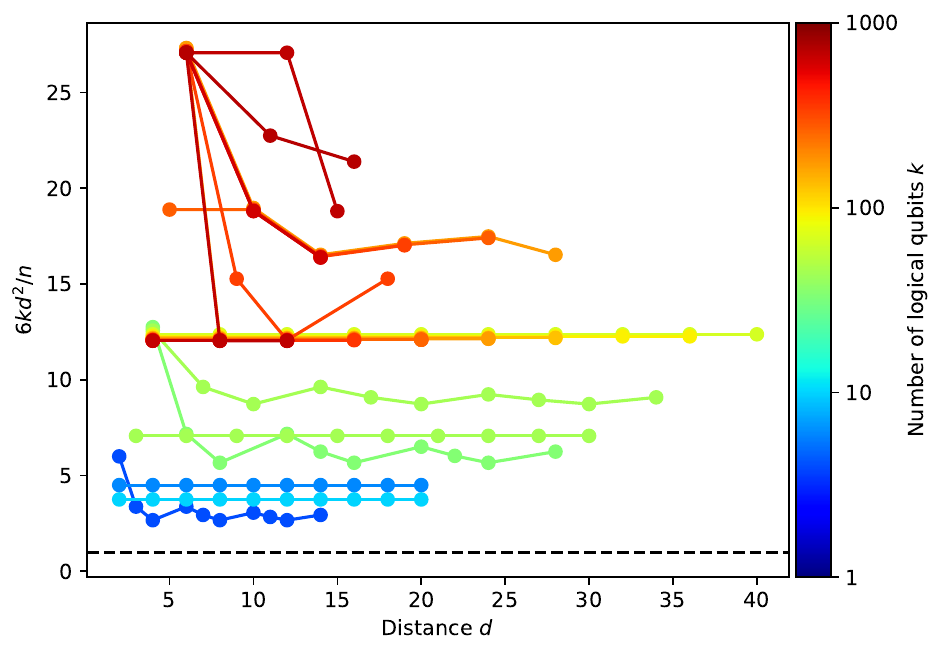}
    \caption{Parameters of semi-hyperbolic Floquet codes. Each line corresponds to a family of semi-hyperbolic codes constructed by fine-graining an 8.8.8 colour code tiling, with $l$ increasing by one from left to right (starting at $l=1$). We plot $6kd^2/n$ on the $y$-axis against $d$ on the $x$-axis (where $d$ is the embedded distance of the Floquet code). Note that $6kd^2/n=1$ for the planar Floquet code (black dashed line) and hence this ratio corresponds to the multiplicative saving in the number of physical qubits relative to the planar Floquet code to obtain a fixed target $k$ and $d$.}
    \label{fig:semi_hyperbolic_families_nk2n}
\end{figure}

We define a family of semi-hyperbolic Floquet codes from a hyperbolic colour code tiling $\mathcal{T}$ and the fine-graining parameter $l$ to obtain a semi-hyperbolic tiling~$\mathcal{T}_l$ (see \Cref{sec:semi_hyperbolic_tilings}), which we use to define the semi-hyperbolic Floquet code $\mathcal{F}(\mathcal{T}_l)$.
Let us denote the parameters of the Floquet code $\mathcal{F}(\mathcal{T})$ derived from $\mathcal{T}$ by $[[n,k,d]]$ and we denote the parameters of $\mathcal{F}(\mathcal{T}_l)$ by $[[n_l,k_l,d_l]]$.
The topology of the surface is unchanged by the fine-graining procedure so we have $k_l=k$ and from \Cref{eq:semi_hyperbolic_num_vertices} we have that $n_l=l^2n$.
The minimum length of a homologically non-trivial cycle or co-cycle in the restricted lattices of $\mathcal{T}_l$ will increase by a factor proportional to $l$, and so we have $d_l\geq C_1ld$ for some constant $C_1$.
This leads to parameters $[[n_l,k_l,d_l]]$ of the semi-hyperbolic Floquet code $\mathcal{F}(\mathcal{T}_l)$ satisfying $k_ld_l^2/n_l\geq C_1^2kd^2/n$.
If we fix the tiling $\mathcal{T}$ from which we define a family of semi-hyperbolic Floquet codes $\mathcal{F}(\mathcal{T}_l)$ then we have $k_ld_l^2/n_l\geq C_2$ where $C_2=C_1^2kd^2/n$ is a constant determined by $\mathcal{T}$.
Although a family of semi-hyperbolic Floquet codes defined this way does not have an \textit{asymptotic} improvement over the parameters of the honeycomb code, it is still possible to obtain a constant factor improvement if $C_2>1/6$.

\begin{figure}
    \centering
    \includegraphics[width=0.99\linewidth]{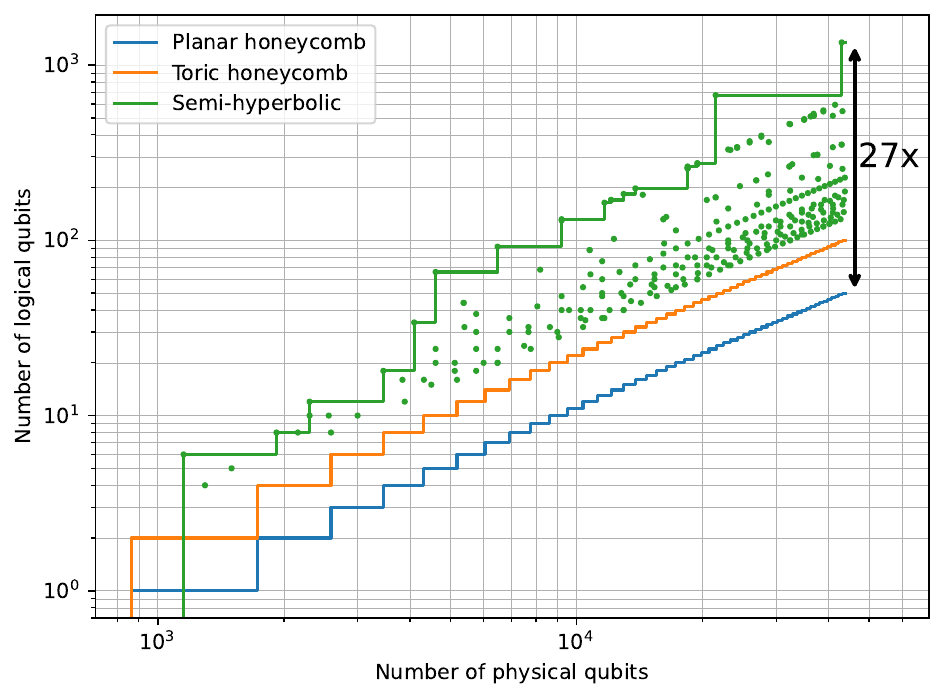}
    \caption{The number of logical qubits that can be encoded with embedded distance $d$ at least 12 using multiple copies of a semi-hyperbolic, toric honeycomb or planar honeycomb floquet code. The EM3 distance is the distance of the code for the EM3 noise model, for which two-qubit Pauli operators can be measured directly without the need for ancillas. Each green circle corresponds to multiple copies of a semi-hyperbolic floquet code, and the green line is the Pareto frontier for the copies of semi-hyperbolic codes considered.}
    \label{fig:n_vs_k_up_to_d12_n44000}
\end{figure}

From our analysis of the parameters of the semi-hyperbolic Floquet codes we have constructed, we find that this constant factor improvement over the planar honeycomb code can be substantial.
In \Cref{fig:semi_hyperbolic_families_nk2n} we show the parameters of families of semi-hyperbolic Floquet codes derived from 8.8.8 colour code tilings.
For some families of semi-hyperbolic codes we have $C_1=1$, i.e.~the line is horizontal on the plot, at least for the system sizes we consider.
For these semi-hyperbolic Floquet code families, we can increase the distance by a factor of $l$ using exactly $l^2\times$ more physical qubits.
Many of these families of semi-hyperbolic codes retain an order-of-magnitude reduction in qubit overhead relative to the planar honeycomb code even for very large distances.

In \Cref{fig:n_vs_k_up_to_d12_n44000} we plot the number of logical qubits that can be encoded (with embedded distance at least 12) as a function of the number of physical qubits available, using multiple copies of semi-hyperbolic Floquet codes and honeycomb codes.
We can encode up to $27\times$ more logical qubits using semi-hyperbolic Floquet codes relative to planar honeycomb codes, including a $>10\times$ increase in the encoding rate for smaller system sizes using fewer than $10,000$ physical qubits.

\subsection{Simulations}\label{sec:simulations}

So far we have focused our analysis on code parameters, however logical error rate performance for realistic noise models is more relevant to understand the practical utility of the constructions.
In particular, hyperbolic Floquet codes derived from an $r.g.b$ tiling have higher plaquette stabiliser weights $2r$, $2g$ and $2b$, and each detector is formed from the parity of $2r$, $2g$ or $2b$ check operator measurements.
We might be concerned that these higher stabiliser weights lead to worse logical error rate performance, despite their improved code parameters.
However, by performing numerical simulations, we show in this section that hyperbolic Floquet codes have a logical error rate performance that is comparable to that of the honeycomb code even at high physical error rates, but with significantly reduced resource overheads.

Our numerical simulations compare the performance of semi-hyperbolic Floquet codes with planar honeycomb codes (for the EM3 and SD6 noise models) as well as rotated planar surface codes (for the SD6 noise models).
All simulations were carried out using Stim~\cite{Gidney2021stimfaststabilizer} to simulate the circuits and construct matching graphs and PyMatching~\cite{higgott2023sparse} to decode.
For the simulations of planar honeycomb codes and surface codes, we used the open-source simulation software~\cite{gidney2022generation} written for Ref.~\cite{gidney2022benchmarking}.
In all figures, we define the logical error rate as the probability that \textit{any} of the encoded logical qubits fail.
Note that we have made some Stim circuits of hyperbolic and semi-hyperbolic Floquet codes, as well as tables of their parameters, available on GitHub~\cite{higgott2023data}.

\subsubsection{Thresholds}

\begin{figure}
    \centering
        \subfloat[]{
        \includegraphics[width=0.45\textwidth]{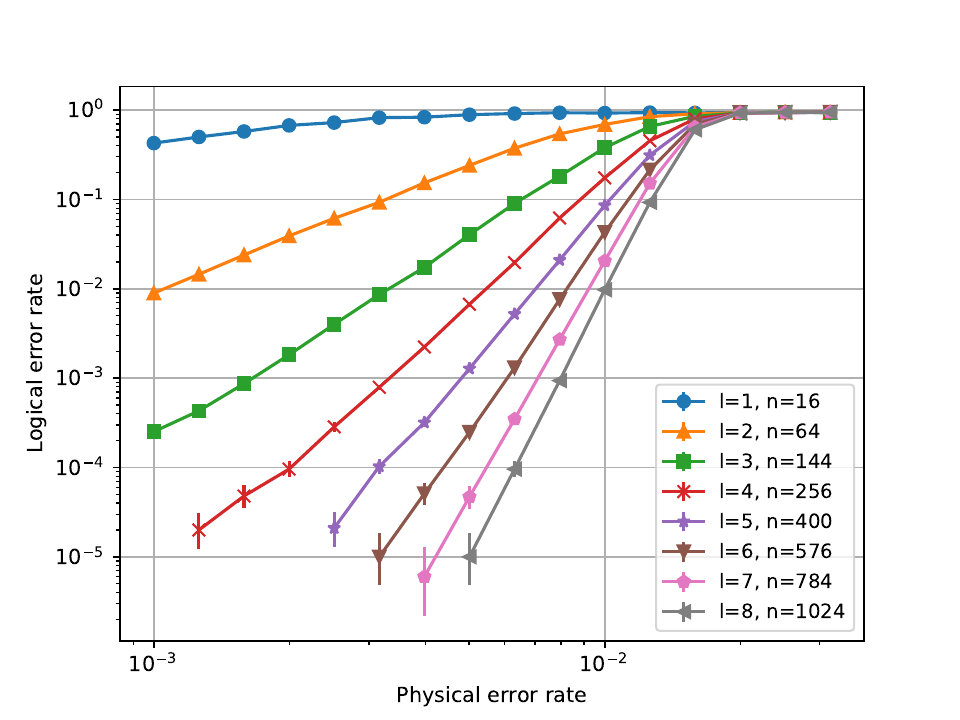}
            \label{subfig:bolza_semi_hyperbolic_threshold}
        }\\
        \subfloat[]{
        \includegraphics[width=0.45\textwidth]{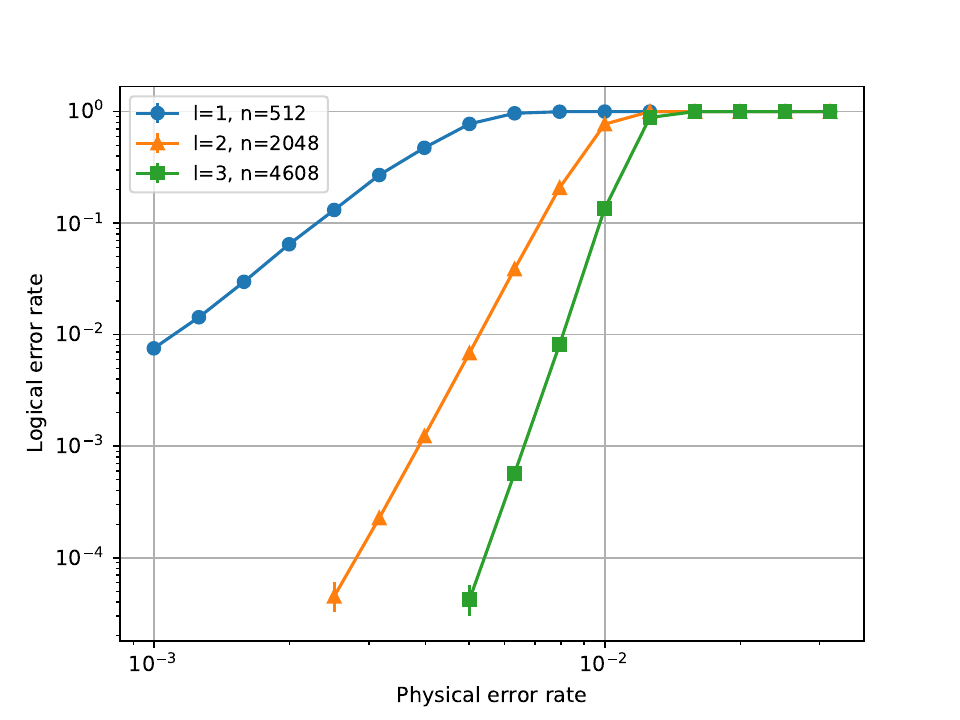}
            \label{subfig:semi_hyperbolic_66_logicals_threshold_em3}
        }
        \caption[Threshold plots of semi-hyperbolic Floquet codes]{Threshold plots of two families of semi-hyperbolic Floquet codes, using an EM3 noise model and 64 rounds. (a) A family of semi-hyperbolic Floquet codes derived from the Bolza surface (see \Cref{tab:bolza_semi_hyperbolic_parameters} for code parameters). (b) A family of semi-hyperbolic Floquet codes derived from an 8.8.8 tiling of a genus-33 closed hyperbolic surface (with code parameters given in \Cref{tab:genus_33_semi_hyperbolic}). For both plots, the legend gives the fine-graining parameter $l$ and number of physical qubits $n$ and we observe a threshold of around 1.5\%-2\%, consistent with the threshold of the honeycomb code~\cite{gidney2021fault,gidney2022benchmarking,paetznick2023performance}.}
    \label{fig:semi_hyperbolic_thresholds}
\end{figure}

We expect families of semi-hyperbolic Floquet codes to have the same threshold as planar and toric honeycomb codes, since the bulk of a semi-hyperbolic Floquet code looks identical to a honeycomb code as we increase the fine-graining parameter $l$.
We demonstrate this numerically for two families of semi-hyperbolic Floquet codes in \Cref{fig:semi_hyperbolic_thresholds}, which each have a threshold of at least 1.5\% to 2\%, consistent with the threshold of the honeycomb code~\cite{gidney2021fault,gidney2022benchmarking,paetznick2023performance}.

\begin{table}[h]
\caption{Parameters of a family of semi-hyperbolic floquet codes derived from the Bolza surface. Here, $n_{tot}$ is the total number of physical qubits (including ancillas) for the SD6 noise model. We verified that the EM3 distance $d_e$ equals the embedded distance $d$ reported here for all codes in the table.}
\label{tab:bolza_semi_hyperbolic_parameters}
\centering
\begin{tabular}[t]{|Sc|Sc|Sc|Sc|Sc|Sc|Sc|Sc|}
\hline
$l$ & $k$ & $n$ & $d$ & $kd^2/n$ & $n_{tot}$ & $d_s$ & $kd_s^2/n_{tot}$ \\
\hline
 1 &    4 &     16 &  2 & 1.00 &    40 &  2 & 0.40 \\
 2 &    4 &     64 &  3 & 0.56 &   160 &  4 & 0.40 \\
 3 &    4 &    144 &  4 & 0.44 &   360 &  6 & 0.40 \\
 4 &    4 &    256 &  6 & 0.56 &   640 &  8 & 0.40 \\
 5 &    4 &    400 &  7 & 0.49 & 1,000 & 10 & 0.40 \\
 6 &    4 &    576 &  8 & 0.44 & 1,440 & 12 & 0.40 \\
 7 &    4 &    784 & 10 & 0.51 & 1,960 & 14 & 0.40 \\
 8 &    4 &  1,024 & 11 & 0.47 & 2,560 & 16 & 0.40 \\
 9 &    4 &  1,296 & 12 & 0.44 & 3,240 & 18 & 0.40 \\
10 &    4 &  1,600 & 14 & 0.49 & 4,000 & 20 & 0.40 \\
\hline
\end{tabular}
\end{table}

\begin{table}
\caption{Parameters of a family of semi-hyperbolic floquet codes derived from an 8.8.8 tiling of a genus-33 closed hyperbolic surface.}
\label{tab:genus_33_semi_hyperbolic}
\centering
\begin{tabular}[t]{|Sc|Sc|Sc|Sc|Sc|}
\hline
$l$ & $k$ & $n$ & $d$ & $kd^2/n$ \\
\hline
 1 &   66 &    512 &  4 & 2.06 \\
 2 &   66 &  2,048 &  8 & 2.06 \\
 3 &   66 &  4,608 & 12 & 2.06 \\
 4 &   66 &  8,192 & 16 & 2.06 \\
 5 &   66 & 12,800 & 20 & 2.06 \\
 6 &   66 & 18,432 & 24 & 2.06 \\
 7 &   66 & 25,088 & 28 & 2.06 \\
 8 &   66 & 32,768 & 32 & 2.06 \\
 9 &   66 & 41,472 & 36 & 2.06 \\
10 &   66 & 51,200 & 40 & 2.06 \\
\hline
\end{tabular}
\end{table}

\subsubsection{Overhead reduction}

\begin{figure}
    \centering
    \subfloat[]{
        \includegraphics[width=0.45\textwidth]{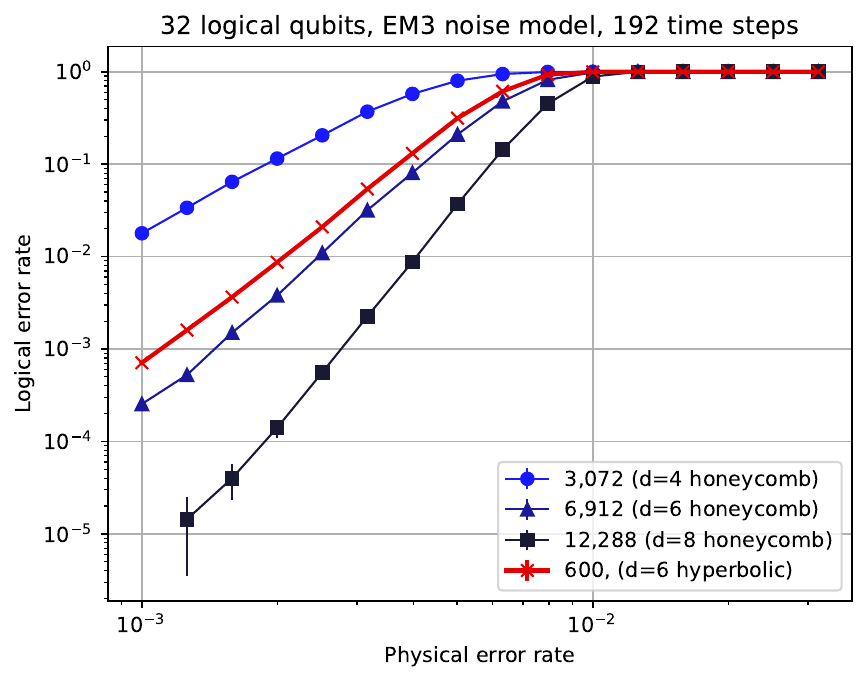}
        \label{subfig:planar_vs_hyperbolic_32_qubits_em3}
        }
        \\
    \subfloat[]{
        \includegraphics[width=0.45\textwidth]{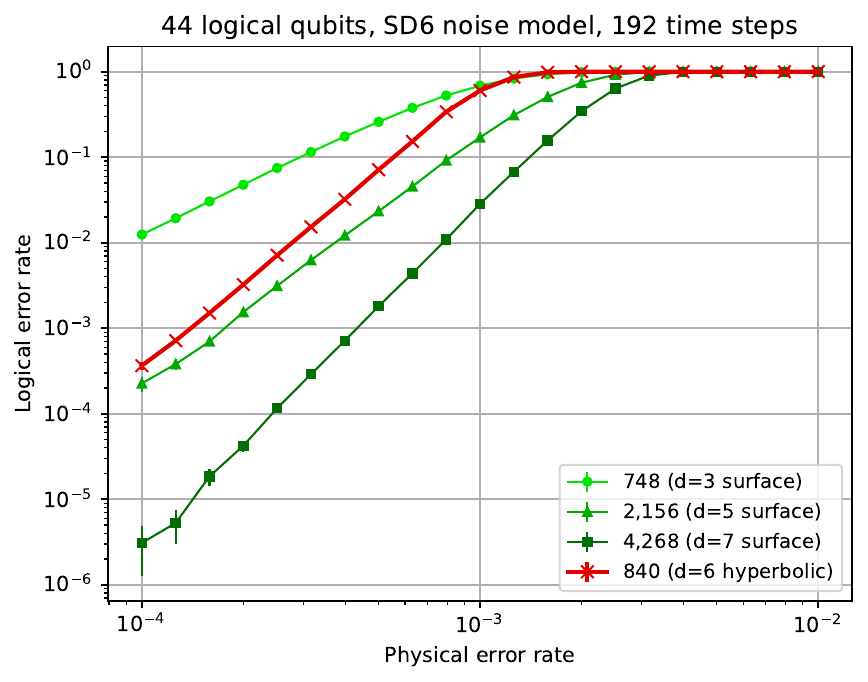}
        \label{subfig:planar_vs_hyperbolic_44_qubits_sd6}
    }
    \caption{(a) Performance of a $[[600,32,d_e=6]]$ hyperbolic Floquet code (derived from a 4.10.10 tiling) compared to 32 copies of planar honeycomb codes, for an EM3 noise model (direct pair measurements). For the honeycomb codes we plot $1-(1-p_{log})^{32}$ where $p_{log}$ is the logical error rate for a single patch (one logical qubit). (b) Performance of a $[[336,44,d_s=6]]$ hyperbolic Floquet code (from an 8.8.8 tiling) compared to 44 copies of rotated surface codes (i.e.~plotting $1-(1-p_{log})^{44}$), for an SD6 noise model. For both (a) and (b), we simulate 192 time steps and give the average of the logical $X$ and $Z$ error rates on the $y$-axis. The legend gives the total number of physical qubits (including ancillas for the SD6 noise model).}
    \label{fig:planar_vs_hyperbolic_medium_sized}
\end{figure}

In this section we study the reduction in qubit overhead that can be achieved using hyperbolic Floquet codes below threshold, for both EM3 and SD6 noise models.

\Cref{fig:planar_vs_hyperbolic_medium_sized} analyses the performance of hyperbolic Floquet codes requiring only a few hundred physical qubits.
In \Cref{subfig:planar_vs_hyperbolic_32_qubits_em3} we simulate the performance of a $[[600,32,d_e=6]]$ hyperbolic Floquet code derived from a 4.10.10 tiling compared to 32 copies of planar honeycomb codes for the Majorana-inspired EM3 noise model, which assumes direct pair measurements.
Achieving the same performance as the hyperbolic Floquet code using planar honeycomb codes requires 6,912 physical qubits ($11.5\times$ more qubits than the hyperbolic Floquet code).
Furthermore, planar honeycomb codes are already known to be $2\times$ to $6\times$ more efficient than surface codes for comparable noise models that assume compilation into direct two-qubit Pauli measurements~\cite{Chao2020optimizationof,paetznick2023performance,gidney2022pair}.

We also show a comparison of a $[[336,44,d_s=6]]$ hyperbolic Floquet code (from an 8.8.8 tiling) in \Cref{subfig:planar_vs_hyperbolic_44_qubits_sd6}, compared to 44 copies of rotated surface codes, for the standard circuit-level depolarising (SD6 noise model).
Including ancillas, the hyperbolic Floquet code requires 840 physical qubits whereas the surface codes require 2156 physical qubits ($2.6\times$ more than the hyperbolic Floquet code).

\begin{figure}
    \centering
    \subfloat[]{
        \includegraphics[width=0.45\textwidth]{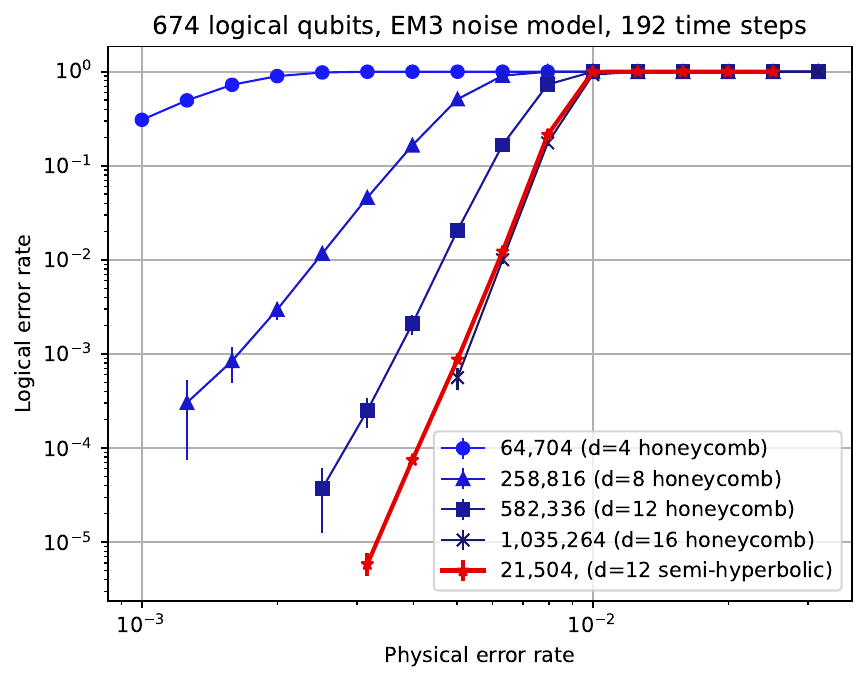}
        \label{subfig:planar_vs_hyperbolic_674_qubits_em3}
        }
    \\
    \subfloat[]{
        \includegraphics[width=0.45\textwidth]{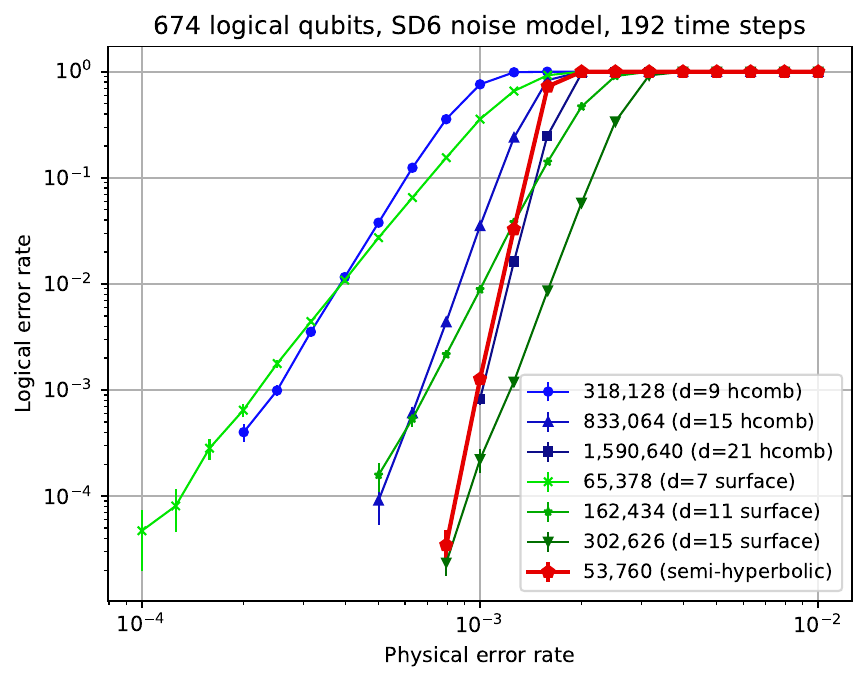}
        \label{subfig:planar_vs_hyperbolic_674_qubits_sd6}
    }
    \caption{Logical error rate vs.~physical error rate for protecting 674 logical qubits using 674 copies of planar honeycomb codes (shades of blue) and a [[21504,674,12]] semi-hyperbolic Floquet code derived from a 8.8.8 tessellation. In (a) an EM3 noise model is used (direct pair measurements without the use of an ancilla) whereas in (b) an SD6 noise model (ancilla-assisted) is used and we also compare with standard surface code circuits (shades of green). For all codes, we simulate 192 time steps and give the average of the logical $X$ and $Z$ error rates on the $y$-axis. For the honeycomb and surface codes we plot $1-(1-p_{log})^{674}$ where $p_{log}$ is the logical error rate for a single patch (one logical qubit). The legend gives the total number of physical qubits (including ancillas for the SD6 noise model).}
    \label{fig:planar_vs_hyperbolic_674_qubits}
\end{figure}

We consider larger examples, which achieve an even larger relative reduction in resources compared to Euclidean codes, in \Cref{fig:planar_vs_hyperbolic_674_qubits}.
In \Cref{subfig:planar_vs_hyperbolic_674_qubits_em3} we consider the EM3 noise model, which assumes noisy direct measurement of two-qubit Pauli operators. 
We simulate the performance of an $l=2$ [[21504,674,12]] semi-hyperbolic Floquet code, which belongs to the family of semi-hyperbolic Floquet codes described in \Cref{tab:semi_hyperbolic_674_logicals_family}.
We compare the logical error rate performance of the [[21504,674,12]] code with honeycomb codes for encoding 674 logical qubits over 192 time steps.
We find that the semi-hyperbolic Floquet code matches the performance of $d=16$ honeycomb codes using $1,035,264$ physical qubits, corresponding to a $48\times$ reduction in qubit overhead.
Therefore, given a platform permitting direct two-qubit Pauli measurements and semi-hyperbolic qubit connectivity (e.g.~using a modular architecture), our results suggest that semi-hyperbolic Floquet codes can require over $100\times$ fewer physical qubits than the surface code circuits from Refs.~\cite{Chao2020optimizationof,paetznick2023performance,gidney2022pair} for an EM3 noise model at around $p=0.1\%$.

Using a least-squares fit of the four left-most data points to extrapolate the red curve in \Cref{subfig:planar_vs_hyperbolic_674_qubits_em3} to lower physical error rates, we estimate that the [[21504,674,12]] semi-hyperbolic Floquet code has a logical error rate of $\approx 1.8\times 10^{-11}$ at $p=0.1\%$ for the EM3 noise model.
This implies a logical error rate per logical qubit of $\approx 2.7\times 10^{-14}$, or $\approx 4.2\times 10^{-16}$ per logical qubit per round.
We therefore project that we can reach well below the ``teraquop regime''~\cite{gidney2021fault} of $10^{-12}$ logical failure rates using only 32 physical qubits per logical qubit.
In contrast, planar honeycomb codes have been shown to require from 600~\cite{gidney2021fault} to 2000~\cite{paetznick2023performance} physical qubits per logical qubit to achieve the teraquop regime using the same noise model in prior work.

\begin{table}
\caption{Parameters of a family of semi-hyperbolic floquet codes encoding 674 logical qubits, constructed by fine-graining an 8.8.8 tiling of a genus-337 hyperbolic surface. Note that $d$ here denotes the embedded distance.}
\label{tab:semi_hyperbolic_674_logicals_family}
\centering
\begin{tabular}[t]{|Sc|Sc|Sc|Sc|Sc|}
\hline
$l$ & $k$ & $n$ & $d$ & $kd^2/n$ \\
\hline
 1 &  674 &  5,376 &  6 & 4.51 \\
 2 &  674 & 21,504 & 12 & 4.51 \\
 3 &  674 & 48,384 & 15 & 3.13 \\
\hline
\end{tabular}
\end{table}

In \Cref{subfig:planar_vs_hyperbolic_674_qubits_sd6}, we instead consider the SD6 noise model, which uses an ancilla qubit to measure each check operator in the presence of standard circuit-level depolarising noise.
We compare a semi-hyperbolic Floquet code (the same as used in \Cref{subfig:planar_vs_hyperbolic_674_qubits_em3}) with honeycomb and surface codes for encoding 674 logical qubits over 192 time steps.
In this noise model, we find that the semi-hyperbolic Floquet code matches the performance of planar honeycomb codes with SD6 distance $d_s=21$ which require $n_{tot}=1,590,640$ qubits, whereas the semi-hyperbolic Floquet code only needs $53,760$ qubits (including ancillas), a $29.6\times$ saving in resources.
At a physical error rate slightly below $0.1\%$, our semi-hyperbolic Floquet code achieves a $5.6\times$ reduction in qubit overhead relative to surface codes, matching the logical error rate of $d=15$ surface codes which use $302,626$ physical qubits.
Note that the curve for the $d=15$ surface code has a shallower gradient than that of the semi-hyperbolic Floquet code, indicating that the semi-hyperbolic Floquet code has a higher SD6 distance.
Therefore, we would expect that the semi-hyperbolic Floquet code will offer a bigger advantage over surface codes at lower physical error rates.
If the semi-hyperbolic Floquet code has SD6 distance $d_s=21$, then it would have similar performance to $d=21$ surface codes at very low physical error rates, which would use $593,794$ physical qubits ($881$ per patch), $11\times$ more than the 53,760 needed for our semi-hyperbolic Floquet code.

\subsection{Small examples}

\begin{figure}
    \centering
    \subfloat[]{
        \includegraphics[width=0.45\textwidth]{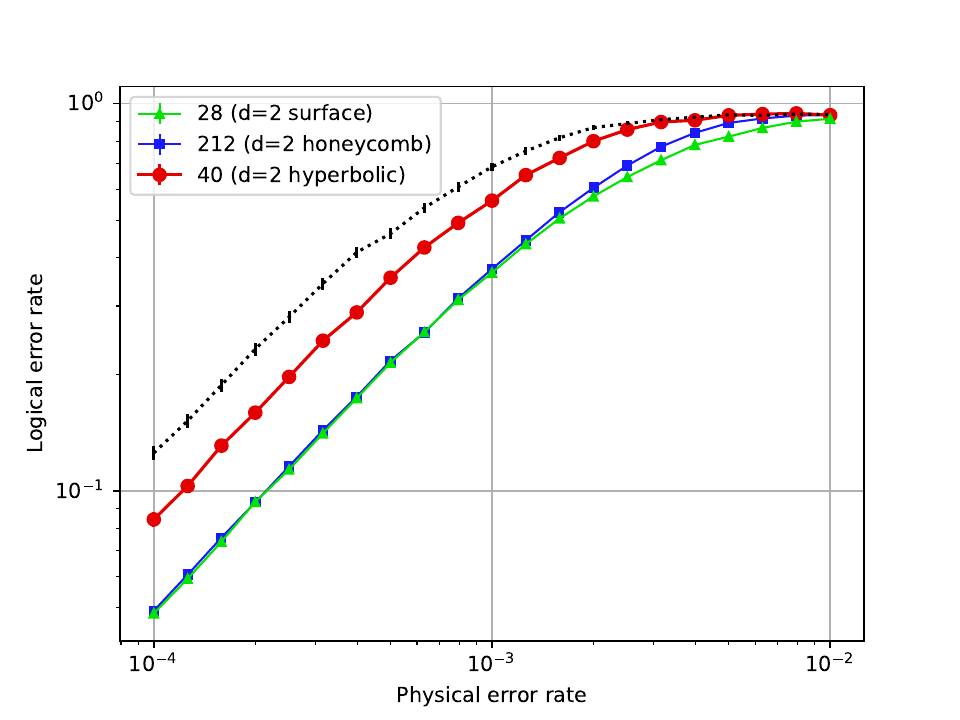}
        \label{subfig:hyperbolic_40_qubit_sd6}
    }\\
    \subfloat[]{
        \includegraphics[width=0.45\textwidth]{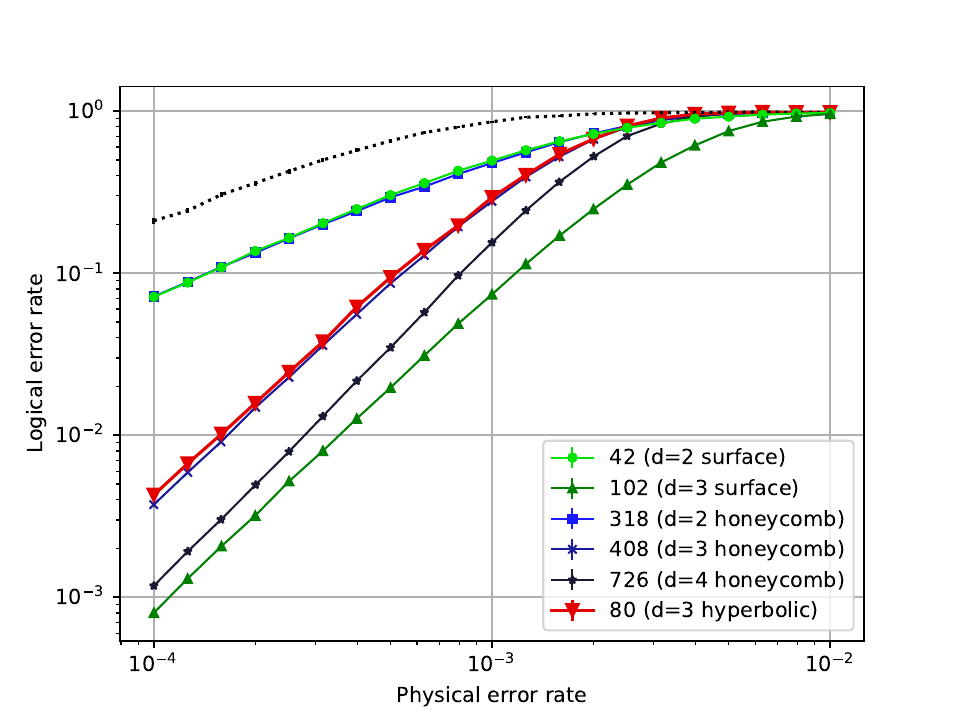}
        \label{subfig:hyperbolic_80_qubit_sd6}
    }
    \caption{Logical error rates vs.~physical error rates for encoding $k$ logical qubits using $k$ copies of small planar honeycomb or surface codes, compared to a hyperbolic floquet code encoding $k$ logical qubits, for 96 time steps using a standard depolarising (SD6) noise model. 
    In (a) we have $k=4$ for the Bolza Floquet code, which has SD6 distance $d_s=2$, whereas in (b) we have $k=6$ for a hyperbolic Floquet code with $d_s=3$.
    The hyperbolic Floquet codes in (a) and (b) are both derived from 8.8.8 tilings. The legend gives the total number of qubits, with the SD6 distance given in brackets.}
    \label{fig:small_hyperbolic_vs_planar_sd6}
\end{figure}

We also study the performance of small hyperbolic and semi-hyperbolic Floquet codes that might be amenable to experimental realisation in the near future.
In \Cref{fig:small_hyperbolic_vs_planar_sd6} we show the performance of a small hyperbolic Floquet codes (with SD6 distance 2 and 3) for an SD6 noise model, compared to small planar honeycomb and surface codes.
To achieve a given logical performance, the hyperbolic Floquet codes in both \Cref{subfig:hyperbolic_40_qubit_sd6} and \Cref{subfig:hyperbolic_80_qubit_sd6} use around $5\times$ fewer physical qubits than planar honeycomb codes but use a similar number of physical qubits to surface codes.

\begin{figure}
    \centering 
    \subfloat[]{
        \includegraphics[width=0.45\textwidth]{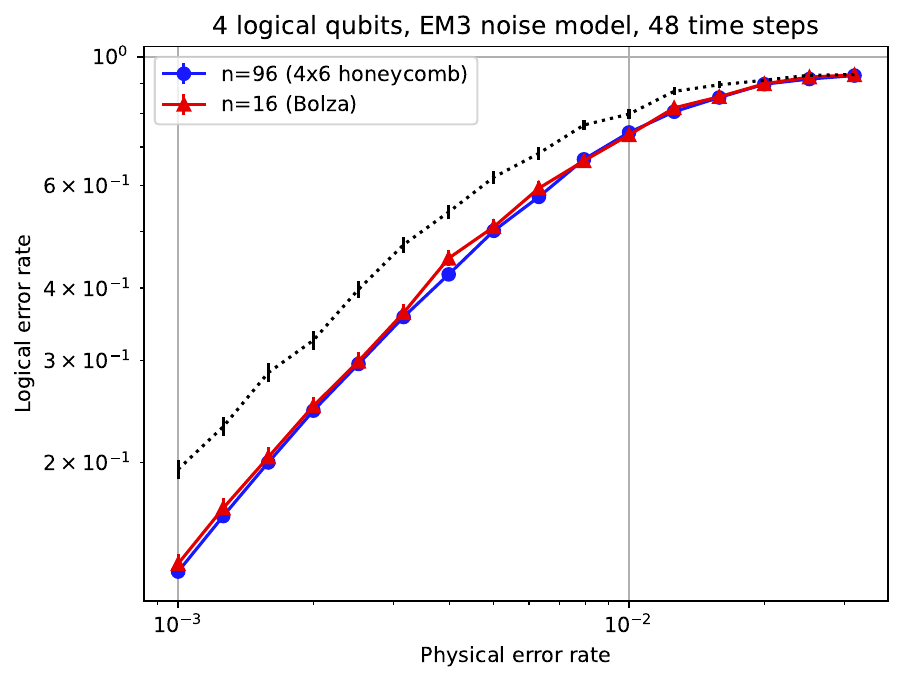}
        \label{subfig:bolza_l1_em3}
    }\\
    \subfloat[]{
        \includegraphics[width=0.48\textwidth]{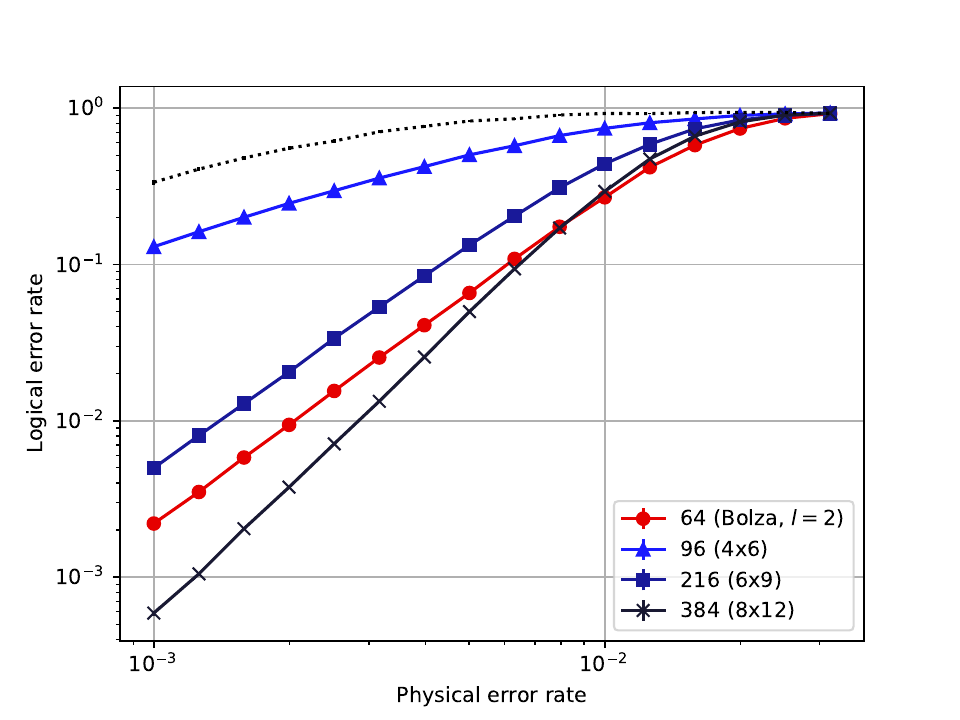}
        \label{subfig:bolza_l2_em3}
    }
    \caption{Logical error rates vs.~physical error rates for encoding 4 logical qubits using 4 copies of small planar honeycomb codes (blue) or a (semi-)hyperbolic floquet code encoding 4 logical qubits (red), for the EM3 noise model (check operators are measured directly without ancillas). In (a) we use a hyperbolic code derived from the Bolza surface that has distance 2 in this noise model, whereas in (b) we use an $l=2$ semi-hyperbolic code derived from the Bolza surface that has distance 3. The legend gives the total number of qubits, with the dimension of the lattice given in parentheses for the planar honeycomb codes. All circuits use 16 rounds of measurements, and the $y$ axis gives the average of the logical $X$ and $Z$ error rates. The black dotted line is the logical failure rate for the (semi-)hyperbolic code for a decoder which always predicts that no logical observable has been flipped. Note that the $4\times 6$, $6\times 9$ and $8\times 12$ planar honeycomb codes have distances 2, 3 and 4 respectively for this EM3 noise model.}
    \label{fig:bolza_em3}
\end{figure}

For the EM3 noise model, we compare to planar honeycomb codes in \Cref{subfig:bolza_l1_em3} and find that the $d=2$ hyperbolic Floquet code derived from the Bolza surface (using 16 qubits) has identical performance to four copies of a $d=2$ planar honeycomb code using $6\times$ more physical qubits.
We also study the performance (again for the EM3 noise model) of an $l=2$ semi-hyperbolic Floquet code derived from the Bolza surface in \Cref{subfig:bolza_l2_em3} and find it to be $3.4\times$ more efficient than four copies of $d=3$ planar honeycomb codes achieving a similar logical error rate.

\section{Conclusion}
\label{sec:conclusion}

In this work, we have constructed Floquet codes derived from colour code tilings of closed hyperbolic surfaces.
These constructions include hyperbolic Floquet codes, obtained from hyperbolic tilings, as well as semi-hyperbolic Floquet codes, which are derived from hyperbolic tilings via a fine-graining procedure.
We have given explicit examples of hyperbolic Floquet codes with improved encoding rates relative to honeycomb codes and have shown how semi-hyperbolic Floquet codes can retain this advantage while enabling improved $\sqrt{n}$ distance scaling.

We have used numerical simulations to analyse the performance of our constructions for two noise models: a Majorana-inspired `EM3' noise model, which assumes direct noisy two-qubit measurements, and a standard circuit-level depolarising `SD6' noise model, which uses ancilla qubits to assist the measurement of check operators.
For the EM3 noise model, we compare a semi-hyperbolic Floquet code that encodes 674 logical qubits into 21,504 physical qubits with 674 copies of planar honeycomb codes, and find that the semi-hyperbolic Floquet code uses $48\times$ fewer physical qubits at physical error rates as high as $0.3\%$ to $1\%$.
We demonstrate that this semi-hyperbolic Floquet code can achieve logical error rates below $10^{-12}$ at $0.1\%$ EM3 noise using as few as 32 physical qubits per logical qubit.
This is a significant improvement over planar honeycomb codes, that require 600 to 2000 physical qubits per logical qubit in the same regime~\cite{gidney2021fault,paetznick2023performance}.
For the SD6 noise model at a noise strength of $0.1\%$, we show that the same semi-hyperbolic Floquet code uses around $30\times$ fewer physical qubits than planar honeycomb codes, and around $5.6\times$ fewer qubits than surface codes.
We also construct several small examples of hyperbolic and semi-hyperbolic Floquet codes amenable to near-term experiments, including a 16-qubit hyperbolic Floquet code derived from the Bolza surface.
These small instances are around $3\times$ to $6\times$ more efficient than planar honeycomb codes and are comparable to standard surface codes in the SD6 noise model.

An interesting avenue of future research might be to study the performance of other recent variants and generalisations of Floquet codes~\cite{davydova2023floquet,kesselring2022anyon,aasen2022adiabatic,zhang2023x,bauer2024topological,sullivan2023floquet,davydova2023quantum,townsendteague2023floquetifying,dua2023engineering}.
For example, the CSS Floquet code introduced in Refs.~\cite{davydova2023floquet,kesselring2022anyon} is also defined from any colour code tiling, but with a modified choice of two-qubit check operators.
CSS Floquet codes can therefore also be constructed from the hyperbolic and semi-hyperbolic colour code tilings in \Cref{sec:colour_code_tilings} and, since they have the same embedded homological codes as the Floquet codes we study here, we would expect them to offer similar advantages over CSS honeycomb codes.

Another possible research direction would be to study and design more concrete realisations of hyperbolic and semi-hyperbolic Floquet codes in modular architectures.
This could include studying protocols for realising these modular architectures where links between modules are more noisy or slow~\cite{fowler2010surace,nickerson2013topological,nickerson2014freely,ramette2023faulttolerant}, motivated by physical systems that can be used to realise them, such as ions~\cite{monroe2014large,stephenson2020high,pino2021demonstration}, atoms~\cite{reiserer2015cavity, bluvstein2022quantum} or superconducting qubits~\cite{sahu2022quantum,tu2022high,delaney2022superconducting,Zhu2020waveguide,Imany2022quantum}.

Finally, we have focused here on error corrected quantum \textit{memories}, but more work is needed to demonstrate how our constructions can be used to save resources within the context of a quantum computation.
We note that the Floquet code schedule applies a logical Hadamard to all logical qubits in the code every three sub-rounds; however, to be useful in a computation we must either be able to read and write logical qubits to the memory or address logical qubits individually with a universal gate set.
We expect that reading and writing of logical qubits to and from honeycomb codes can be achieved using Dehn twists and lattice surgery by adapting the methods developed in Ref.~\cite{breuckmann2017semihyperbolic} for hyperbolic surface codes to (semi-)hyperbolic Floquet codes, however a more detailed analysis is required.
Further work could also investigate (semi-)hyperbolic Floquet codes that admit additional transversal logical operations~\cite{quintavalle2022partitioning,webster2023transversal} such as fold-transversal logical gates~\cite{breuckmann2022foldtransversal}, examples of which have been demonstrated for hyperbolic surface codes.

\section*{Acknowledgements}

OH would like to thank Craig Gidney for helpful discussions, including suggesting \Cref{fig:n_vs_k_up_to_d12_n44000}, as well as Barbara Terhal and Matt McEwen for discussions that improved the section on biplanar and modular architectures.
OH acknowledges support from the Engineering
and Physical Sciences Research Council [grant number EP/L015242/1] and a Google PhD fellowship.

\bibliographystyle{unsrt}
\bibliography{references}

\appendix

\section{Additional results}

\begin{figure}[h]
    \centering
    \subfloat[]{
        \includegraphics[width=0.45\textwidth]{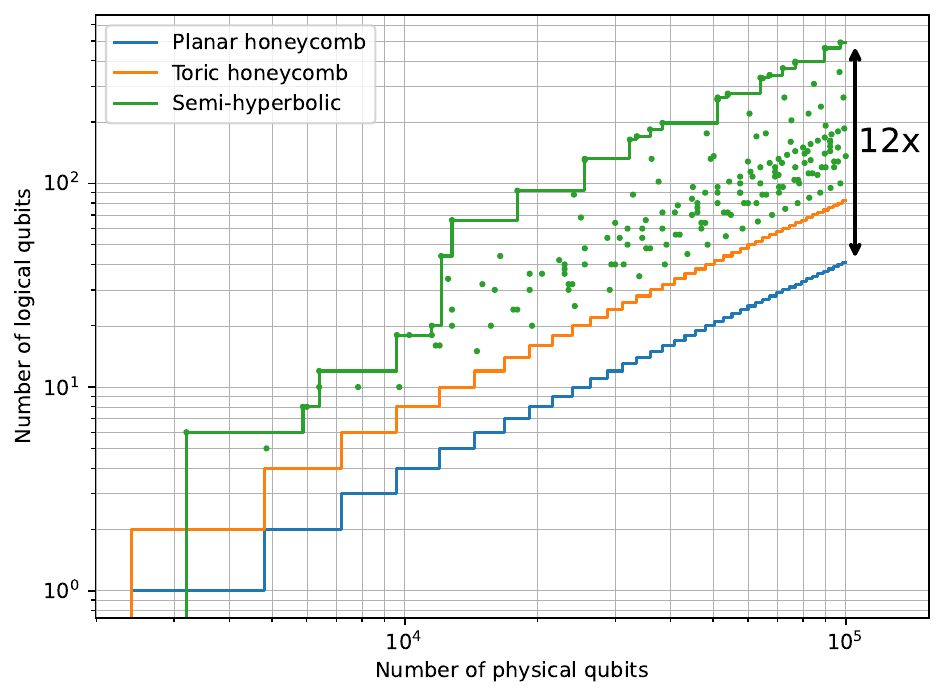}
        \label{subfig:n_vs_k_floquet_d_20}
    }\\
    \subfloat[]{
        \includegraphics[width=0.45\textwidth]{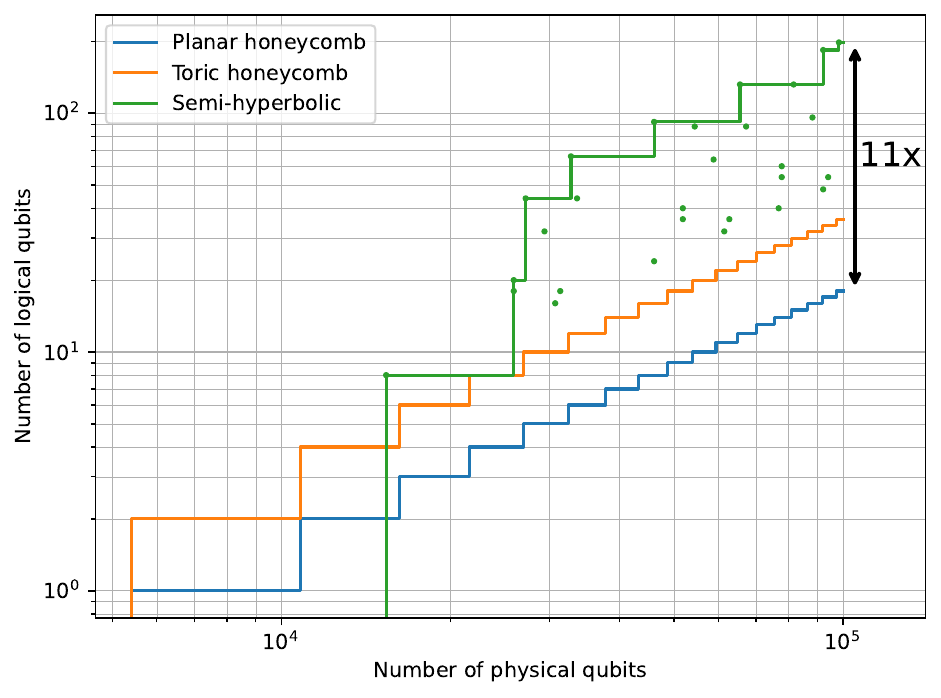}
        \label{subfig:n_vs_k_floquet_d_30}
    }
    \caption{The number of logical qubits that can be encoded using multiple copies of a semi-hyperbolic, toric honeycomb or planar honeycomb floquet code. In (a) the distance is required to be at least 20 for an EM3 noise model, whereas in (b) the distance is at least 30.}
    \label{fig:n_vs_k_floquet_d20_d30}
\end{figure}

\Cref{fig:n_vs_k_floquet_d20_d30} gives the number of logical qubits that can be encoded with embedded distance at least 20 and 30 using semi-hyperbolic Floquet codes or honeycomb codes (these plots are slight variants of \Cref{fig:n_vs_k_up_to_d12_n44000}).

\end{document}